\documentclass[]{JHEP3}
\pdfoutput = 1
\usepackage{graphicx}
\usepackage{amsmath,amssymb,amscd,amsfonts,bm}

\newcommand{\ME}{M}

\newcommand{\as}{\alpha_{\mathrm{s}}}

\newcommand{\LA}{\mathrm{A}}
\newcommand{\LB}{\mathrm{B}}

\newcommand{\La}{\mathrm{a}}
\newcommand{\Lb}{\mathrm{b}}
\newcommand{\Lc}{\mathrm{c}}

\def\ket#1{\big|{#1}\big\rangle}
\def\bra#1{\big\langle{#1}\big|}
\def\brax#1{\big\langle{#1}}   

\def\sket#1{\big|{#1}\big)}
\def\sbra#1{\big({#1}\big|}
\def\sbrax#1{\big({#1}}        

\def\dualL{\raisebox{-5 pt}{$\scriptstyle D$}\!}   

\newbox\charbox
\newbox\slabox
\def\s#1{{      
        \setbox\charbox=\hbox{$#1$}
        \setbox\slabox=\hbox{$/$}
        \dimen\charbox=\ht\slabox
        \advance\dimen\charbox by -\dp\slabox
        \advance\dimen\charbox by -\ht\charbox
        \advance\dimen\charbox by \dp\charbox
        \divide\dimen\charbox by 2
        \raise-\dimen\charbox\hbox to \wd\charbox{\hss/\hss}
        \llap{$#1$}
}}

\title{Parton showers with quantum interference: leading color, with spin}

\author{Zolt\'an Nagy \\
Theory Group,
CERN\\
CH-1211 Geneva 23, Switzerland\\
E-mail: \email{Zoltan.Nagy@cern.ch}
}

\author{Davison E. Soper\\
Institute of Theoretical Science\\
University of Oregon\\
Eugene, OR  97403-5203, USA\\
{\rm and}\\
Theory Group,
CERN\\
CH-1211 Geneva 23, Switzerland\\
E-mail: \email{soper@uoregon.edu}
}

\abstract{
We have previously described a mathematical formulation for a parton shower based on the approximation of strongly ordered virtualities of successive parton splittings. Quantum interference, including interference among different color and spin states, was included. A practical numerical implementation strategy was left unspecified. In a subsequent paper, we showed that if we add the further approximations of taking only the leading color limit and averaging over spins, we obtain a shower evolution that can be implemented as a Markov process. In this paper, we outline a strategy for including the correlations induced by parton spins.
}

\preprint{CERN-PH-TH-2008-087}

\begin{document}


\section{Introduction}

In Ref.~\cite{NSshower}, we described how the evolution of a leading order parton shower could be formulated so that the shower is built on the approximation that the virtuality in each successive parton splitting is much smaller than the virtuality in the preceding splitting. No further approximation is made. The evolution equations thus necessarily account for interference graphs, for the color state of the partons, and for the spin carried by each parton. The evolution equations generate a desired cross section as a nested set of integrals that could, in principle, be performed by numerical integration. However, we did not present a method for implementing the integrations in a practical manner.

Within the general framework of Ref.~\cite{NSshower}, it is possible to make further approximations. In particular, one can take the leading color limit, $1/N_\Lc^2 \to 0$, where $N_\Lc = 3$ is the number of colors. This greatly simplifies the structure of the evolution. Furthermore, at each splitting one can average over the spin of the mother parton and sum over the spins of the daughter partons. Both of these approximations are commonly used in parton shower Monte Carlo event generators. Here, we still keep interference graphs, as in those event generators that are based on color dipoles. In Ref.~\cite{NSspinless}, we saw that the general formalism with these further approximations can be formulated in a standard style of calculation as a Markov process.

What might be the effect of averaging over spins? Partons carry quantum spins and evolution creates multi-parton states in which the spins of the partons are correlated with one another. At each successive splitting, the spin of the mother parton can affect the distribution of the momenta of the daughter partons. In this paper, we assume that the spins of the final partons that appear at the end of the shower are not measured. That is, we take the square of the amplitude to produce each spin state and we sum this over the spins. Nevertheless, the spin distribution of the intermediate partons is imprinted on the momentum distribution of the partons. Thus, to get the momentum distribution right, one needs to follow the spins.

Additionally, parton spin correlations are important for particles that decay via their coupling to $W$-bosons and, in general, for beyond-the-standard-model particles, as described in Ref.~\cite{HerwigSpin}. For this reason also, one wants to include spin effects in a QCD parton shower in a simple way that can be extended to other interactions. Finally, perturbative QCD matrix elements contain the full spin information for the intermediate virtual particles. If one wants a shower to match these matrix elements in the soft and collinear limits, then the shower should also contain the full spin information for the intermediate partons.

The purpose of this paper is to study how, still working in the leading color approximation, one can put back the effect of the parton spin states on the final momentum distribution. Suppose that one starts with a spin averaged shower. Then there is a weight factor that should be associated with each event that gives the probability to get this event with spin included divided by the probability to get this event (with the same splitting history) in a spin averaged shower. Our approach will be to calculate the weight factor and associate it with the event, so that the cross section to produce a given configuration of partons is proportional to the sum of the weights of the events for which the partons are in that configuration. 

It might seem that a calculation of the weight factor is very complicated, involving, as it does, the entangled spins of states with many partons. However, following an insight in a work by J.~Collins \cite{JCCspin}, we find that the calculation of the spin weight factor is quite straightforward and uses an amount of computational resources, both memory and time, that is proportional to the number of partons in the event.

The approach of this paper can be compared to that of the two most commonly used parton shower event generators. In \textsc{Pythia} \cite{Pythia}, one simply averages over spins. However, some spin induced correlations are incorporated by letting the azimuthal angle of each splitting be correlated with azimuthal angle of the splitting that produced the mother parton, following the prescription of Ref.~\cite{WebberCorrelations}. In \textsc{Herwig} \cite{Herwig}, spin is included in a fashion that produces the spin correlations in the limit that the splittings are collinear or soft$\times$collinear. This leaves out purely soft splittings, for which the angular ordering prescription of \textsc{Herwig} is approximate with respect to azimuthal angle correlations even after averaging over spin. The method \cite{JCCspin, KnowlesSpin, HerwigSpin} has the advantage of not requiring weight functions. However, the method is not consistent with the use of exact momentum conservation at each splitting nor is it compatible with the inclusion of the exact angular correlations arising from soft gluon interference diagrams. Each of these is important in the formulation of Refs.~\cite{NSshower, NSspinless}.

In the following section, we describe how spin appears if we view it as adjoined to a spin averaged shower. This leads to the definition of the spin weight factor. In Sec.~\ref{sec:spineval}, which forms the heart of this paper, we describe how to evaluate the spin weight factor. Then in Sec.~\ref{sec:properties} we investigate some properties of this factor. In Sec.~\ref{sec:example1} we give a numerical example of how one step in the evolution of the spin weight factor works and in Sec.~\ref{sec:3steps} we give a numerical example of how three steps in this evolution work together to correlate azimuthal angles of parton splittings. In Sec.~\ref{sec:Alk}, we turn to a technical topic, how we partition the coherent radiation from two partons, call them $l$ and $k$, into two terms, one treated as radiation from parton $l$ and one treated from parton $k$. Our treatment of the partitioning function $A_{lk}$ generalizes our earlier work on this function, allowing, in particular, an $A_{lk}$ that is spin dependent. We present several possible choices for $A_{lk}$, any of which are compatible with the methods of sections \ref{sec:spinin} through \ref{sec:3steps}. The numerical examples of sections \ref{sec:example1} and \ref{sec:3steps} are based on one of these choices. Finally, Sec.~\ref{sec:conclusions} contains some concluding remarks.

\section{Incorporating spin in the parton shower}
\label{sec:spinin}

In this section we take the general parton shower formalism of Ref.~\cite{NSshower} and make the leading color approximation at each step of the shower but retain the spin information. In Ref.~\cite{NSspinless}, we have seen how to formulate a leading color, spin averaged shower. We write the operators that occur in the leading color shower with spin as products of the functions that apply for the spin-averaged shower with specific operators that act on the partonic spin space. This will leave us with a matrix element in the partonic spin space. This allows us to express the parton shower with spin using the parton shower without spin and then incorporating spin as a weight factor. In the following section, we will see how this spin matrix element can be evaluated.

\subsection{Parton splitting with spin}

We consider a shower in the leading color approximation, but keeping the quantum spins. We start with the complete shower evolution as described in Ref.~\cite{NSshower}, but we make the leading color approximation as described in Ref.~\cite{NSspinless}. We need states representing a set of $m$ final state partons and two initial state partons with momenta $p$ and flavors $f$ in a color state $c$ specified by a color string configuration. Our states are further specified by two spin indices, $s$ and $s'$, for each parton. The spin index $s$ describes the spin of that parton in the quantum amplitude $\ket{M}$ while the spin index $s'$ describes the spin of the parton in the conjugate amplitude $\bra{M}$. We call this state $\sket{\{p,f,c,s',s\}_{m}}$. Our notation is that the partons carry labels $\mathrm{a},\mathrm{b},1,\dots,m$ and that, for instance, $\{p\}_m$ denotes the ordered set of momenta $\{p_\La,p_\Lb,p_1,\cdots,p_m\}$.

The states evolve with a linear operator ${\cal U}^{\rm lc}(t_{\rm f},t')$ that acts on the space for which the $\sket{\{p,f,c,s',s\}_{m}}$ are basis states. Here the superscript ``lc'' designates quantities in the leading color approximation. The evolution parameter is a shower time $t$, the logarithm of the virtuality in parton splittings. The operator ${\cal U}^{\rm lc}(t_{\rm f},t')$ effects the evolution from a time $t'$ to a final time $t_{\rm f}$ at which perturbative shower development is halted. The evolution equation for ${\cal U}^{\rm lc}$ can be specified by stating the action of ${\cal U}^{\rm lc}$ on a general state $\sket{\{p,f,c,s',s\}_{m-1}}$ with $m-1$ final state partons,\footnote{In Refs.~\cite{NSshower} and \cite{NSspinless}, we started with a state of $m$ partons, so that the splitting resulted in a state with $m+1$ final state partons. In this paper it is most convenient to decrease $m$ by one.}
\begin{equation}
\begin{split}
\label{eq:evolutiondetaillc}
{\cal U}^{\rm lc}(t_{\rm f},t')&\sket{\{p,f,c,s',s\}_{m-1}} =
\\ &
\Delta^{(0)}(t_{\rm f},t';\{p,f,c\}_{m-1})\sket{\{p,f,c,s',s\}_{m-1}}
\\
& + 
\int_{t'}^{t_{\rm f}}\! d\tau\ 
\frac{1}{m!}
\int \big[d\{\hat p,\hat f,\hat c\}_{m}\big]
\sum_{\{\hat s',\hat s\}_{m}}
{\cal U}^{\rm lc}(t_{\rm f},\tau)\,
\sket{\{\hat p,\hat f,\hat c,\hat s',\hat s\}_{m}}
\\&\quad \times
\sbra{\{\hat p,\hat f,\hat c,\hat s',\hat s\}_{m}}
{\cal H}^{\rm lc}_{\rm I}(\tau)\sket{\{p,f,c,s',s\}_{m-1}}
\,\Delta^{(0)}(\tau,t';\{p,f,c\}_{m-1})
\;\;.
\end{split}
\end{equation}
The factor $\Delta^{(0)}(\tau,t';\{p,f,c\}_{m})$ is the Sudakov factor that gives the probability for the state not to undergo a parton splitting between shower times $t'$ and $\tau$. The Sudakov factor is the same as in the leading color, spin averaged approximation, which we designate with a superscript $(0)$. Thus the first term in Eq.~(\ref{eq:evolutiondetaillc}) represents the possibility that the system evolves from $t$ to $t_{\rm f}$ with no splitting. In the next term, the state evolves from $t$ to $\tau$ with no splitting, then undergoes a splitting according to the splitting operator ${\cal H}^{\rm lc}_{\rm I}$. After the splitting there are $m$ final state partons, with the new parton carrying the label $m$. We integrate over the time $\tau$. We also integrate and sum over the momenta, flavors, and colors of the new partons after the splitting, as in the shower averaged over spin. Since we now include spin, we also need to sum over the spin indices $\{s',s\}_{m}$ of the partons after the splitting.

Using Refs.~\cite{NSshower} and \cite{NSspinless}, we can write the matrix elements of the splitting operator in the leading color approximation as
\begin{equation}
\label{eq:HLeadingColor}
\begin{split}
\big(\{\hat p,\hat f,{}&\hat c,\hat s',\hat s\}_{m}\big|
{\cal H}^{\rm lc}_{\rm I}(t)\sket{\{p,f,c,s',s\}_{m-1}}
\\={}&
\sum_{l}
m\
\frac
{n_\Lc(a) n_\Lc(b)\,\eta_{\La}\eta_{\Lb}}
{n_\Lc(\hat a) n_\Lc(\hat b)\,
 \hat \eta_{\La}\hat \eta_{\Lb}}\,
\frac{
f_{\hat a/A}(\hat \eta_{\La},\mu^{2}_{F})
f_{\hat b/B}(\hat \eta_{\Lb},\mu^{2}_{F})}
{f_{a/A}(\eta_{\La},\mu^{2}_{F})
f_{b/B}(\eta_{\Lb},\mu^{2}_{F})}
\\
&\times
\sbra{\{\hat p,\hat f\}_{m}}{\cal P}_{l}\sket{\{p,f\}_{m-1}}\,
\delta\!\left(
t - T_l(\{\hat p,\hat f\}_{m})\right)\,
\\
&\times
\biggl\{
\theta(\hat f_{m} = \mathrm{g})\
\sum_{\substack{k\\ k\ne l}}\
\sbra{\{\hat s',\hat s\}_{m}}
{\cal Y}(l,k;\{\hat f,\hat p\}_{m})
\sket{\{s',s\}_{m-1}}\,
\Phi_{lk}(\{\hat p,\hat f\}_{m})
\\ & \qquad \times
\bra{\{\hat c\}_{m}} a^\dagger_{lk}\ket{\{c\}_{m-1}}\,
\\
&\quad+
\theta(\hat f_{m} \ne \mathrm{g})\,
\sbra{\{\hat s',\hat s\}_{m}}
{\cal Y}(l,l;\{\hat f,\hat p\}_{m})
\sket{\{s',s\}_m}\,
\Phi_{ll}(\{\hat p,\hat f\}_{m})
\\&\qquad\times
\Big[
\theta(f_l \in \{q,\bar q\})
\bra{\{\hat c\}_{m}} a^\dagger_{\rm g}(l)\ket{\{c\}_{m-1}}
+
\theta(f_{l} = \mathrm{g})
\bra{\{\hat c\}_{m}} a^\dagger_{q}(l)\ket{\{c\}_{m-1}}
\Big]
\biggr\}
\;\;.
\end{split}
\end{equation}
The first line on the right hand side of this formula contains factors copied directly from Ref.~\cite{NSshower}. There is a sum over the index $l$ of the parton that splits. Then there is a ratio of parton distribution functions, momentum fractions $\hat \eta$, and the numbers of colors $n_{\rm c}$ carried by the partons. This ratio is 1 for a final state splitting but different from 1 for an initial state splitting.\footnote{Initial state splittings are done with backwards evolution. Our notation is that the momenta $p_\La$ and $p_\Lb$ of the initial state partons denote their physical momenta, whereas the flavors $f_\La$ and $f_\Lb$ denote the flavors leaving the hard interaction, which are the opposite of the physical flavors entering the hard interaction.} The next line concerns the relation of the variables $\{\hat p,\hat f\}_{m}$ and $t$ to the variables $\{p,f\}_{m-1}$. For the flavors, this factor vanishes unless $\hat f_{m} + \hat f_l = f_l$, with the evident definition of adding flavors, and it vanishes unless $\hat f_j = f_j$ for the other partons. For an allowed relationship between $\{\hat f\}_{m}$ and $\{f\}_{m-1}$, the flavor factor is 1. There is a similar factor for the momenta. Given the momenta $\{p\}_{m-1}$, the momenta $\{\hat p\}_{m}$ must lie on a certain three dimensional surface specified by the momentum mapping ${\cal R}_l$ defined in Ref.~\cite{NSshower}. The function $\sbra{\{\hat p,\hat f\}_{m}}{\cal P}_{l}\sket{\{p,f\}_{m-1}}$ contains a delta function on this surface. There is also a delta function that defines the shower time $t$ as the logarithm of the virtuality in the splitting,
\begin{equation}
\label{eq:tdef}
T_l(\{\hat p,\hat f\}_{m}) = 
\log\left(\frac{Q_0^2}{|(\hat p_l +(-1)^{\delta_{l,\La} + \delta_{l,\Lb}}
\hat p_{m})^2 - m^2(\hat f_l + \hat f_{m})|}\right)
\;\;.
\end{equation}
Thus if we integrate $\big(\{\hat p,\hat f,\hat c,\hat s',\hat s\}_{m}\big|
{\cal H}^{\rm lc}_{\rm I}(t)\sket{\{p,f,c,s',s\}_{m-1}}$ over $t$ and the momenta $\{\hat p\}_{m}$, we are really integrating over three variables that describe the splitting of parton $l$. 

The most important functions in Eq.~(\ref{eq:HLeadingColor}) are the functions 
${\cal Y}(l,k;\{\hat f,\hat p\}_{m})$, which describe parton splitting with gluon emission. These have the structure
\begin{equation}
\begin{split}
\label{eq:Ylkdef}
{\cal Y}(l,k;\{\hat f,\hat p\}_{m}) ={}& 
\frac{C_{\rm F}}{\Phi_{lk}(\{\hat p,\hat f\}_{m})}\
\biggl\{
{\cal W}(l,l;\{\hat f,\hat p\}_{m})
\\&
- {\cal A}_{lk}(\{\hat p\}_{m})\,
\left[
{\cal W}(l,k;\{\hat f,\hat p\}_{m})
+{\cal W}(k,l;\{\hat f,\hat p\}_{m})
\right]
\biggr\}
\;\;.
\end{split}
\end{equation}
The functions ${\cal W}(l,l;\{\hat f,\hat p\}_{m})$, ${\cal W}(l,k;\{\hat f,\hat p\}_{m})$, and ${\cal W}(k,l;\{\hat f,\hat p\}_{m})$ are operators on the combined parton spin space. Thus we can specify them by giving their matrix elements
\begin{displaymath}
\sbra{\{\hat s',\hat s\}_{m}}
{\cal W}(l,k;\{\hat f,\hat p\}_{m})
\sket{\{s',s\}_m}
\;\;.
\end{displaymath}
These operators are fully defined in Ref.~\cite{NSshower} and there is little point in repeating the definitions here. However, the physical meaning can be appreciated by using some simple pictures. 

The spin matrix element of ${\cal W}(l,l;\{\hat f,\hat p\}_{m})$ is illustrated in Fig.~\ref{fig:Wll}. In the process described by this function, one of the partons, with label $l$, splits in the quantum amplitude to form partons with labels $l$ and $m$. In the complex conjugate amplitude, parton $l$ also splits to form partons $l$ and $m$. For each parton with label $i$ in the amplitude, the spin before splitting is $s_i$ and the spin after splitting is $\hat s_i$. For each parton with label $i$ in the conjugate amplitude, the spin before splitting is $s_i'$ and the spin after splitting is $\hat s_i'$. The function ${\cal W}(l,l;\{\hat f,\hat p\}_{m})$ has a nontrivial dependence on the spins of the active partons $l$ and $m$, a dependence that is taken from the Feynman rules for the splitting amplitudes. There are several other partons, ``1,'' ``2,''  {\it etc}, that are simply spectators.  For a spectator parton, ${\cal W}(l,l;\{\hat f,\hat p\}_{m})$ contains factors $\delta_{s_i,\hat s_i}$ and $\delta_{\hat s_i',s'_i}$ that keep the before and after spins the same.

\FIGURE{
\centerline{\includegraphics[width = 7 cm]{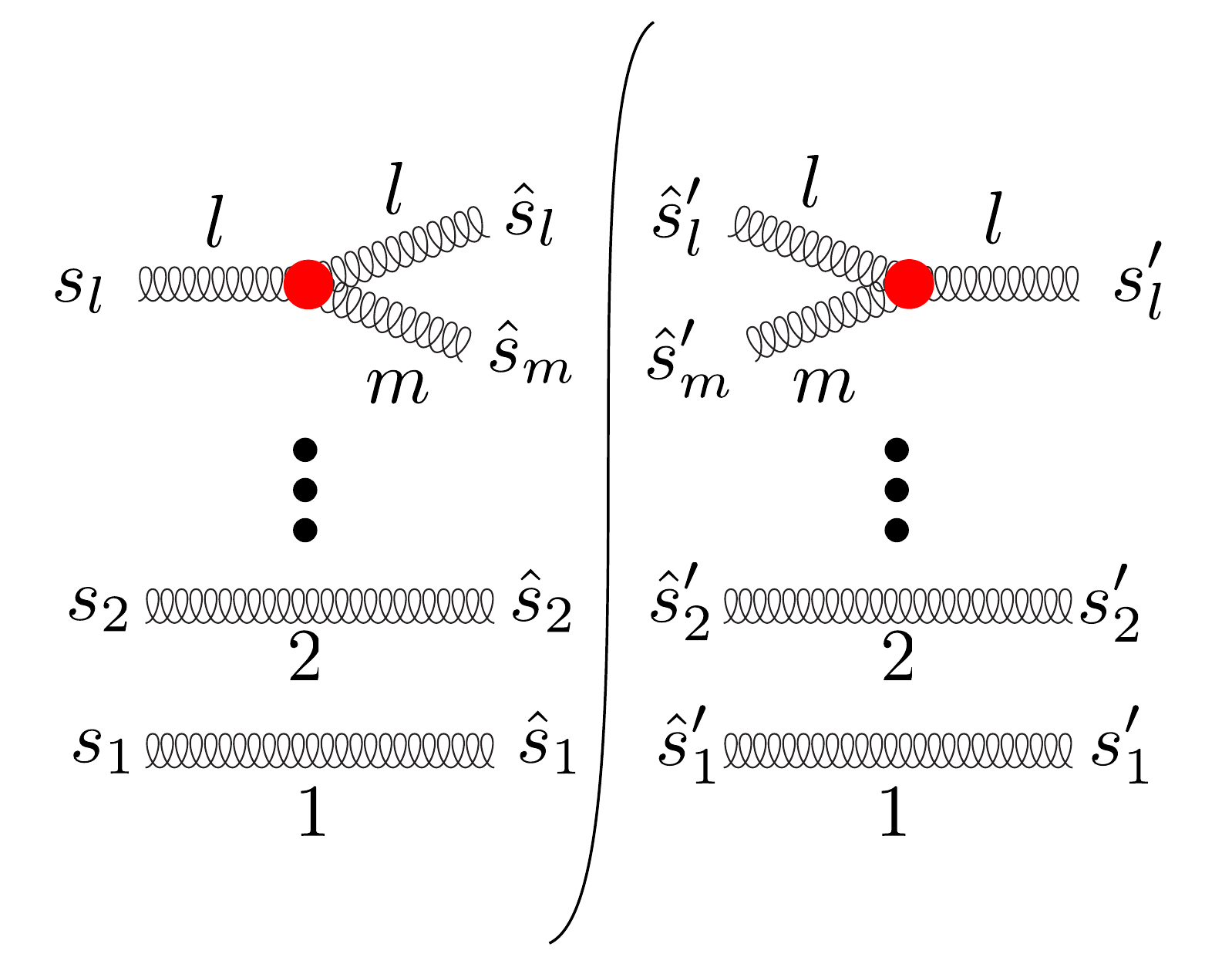}}
\caption{Illustration of the function $W_{ll}$ in Eq.~(\ref{eq:Ylkdef}). Parton $m$ is emitted from parton $l$ in the quantum amplitude and from parton $l$ in the conjugate amplitude.}
\label{fig:Wll}
}

The spin matrix elements of ${\cal W}(l,k;\{\hat f,\hat p\}_{m})$ and ${\cal W}(k,l;\{\hat f,\hat p\}_{m})$ are illustrated in Fig.~\ref{fig:Wlk}. These functions describe quantum interference between two graphs. Consider the function ${\cal W}(l,k;\{\hat f,\hat p\}_{m})$. In the quantum amplitude, parton $l$ splits to form partons with labels $l$ and $m$. In the complex conjugate amplitude, parton $k$ splits to form partons $k$ and $m$. There are several other partons that are simply spectators. The function ${\cal W}(l,k;\{\hat f,\hat p\}_{m})$ is proportional to the unit operator on the spins of partons $l$ and $k$ but has a nontrivial dependence on the spins of parton $m$. This spin dependence is taken from the Feynman rules for the splitting amplitudes making use of the eikonal approximation.

\FIGURE{
\centerline{\hskip 0.3 cm 
\includegraphics[width = 7 cm]{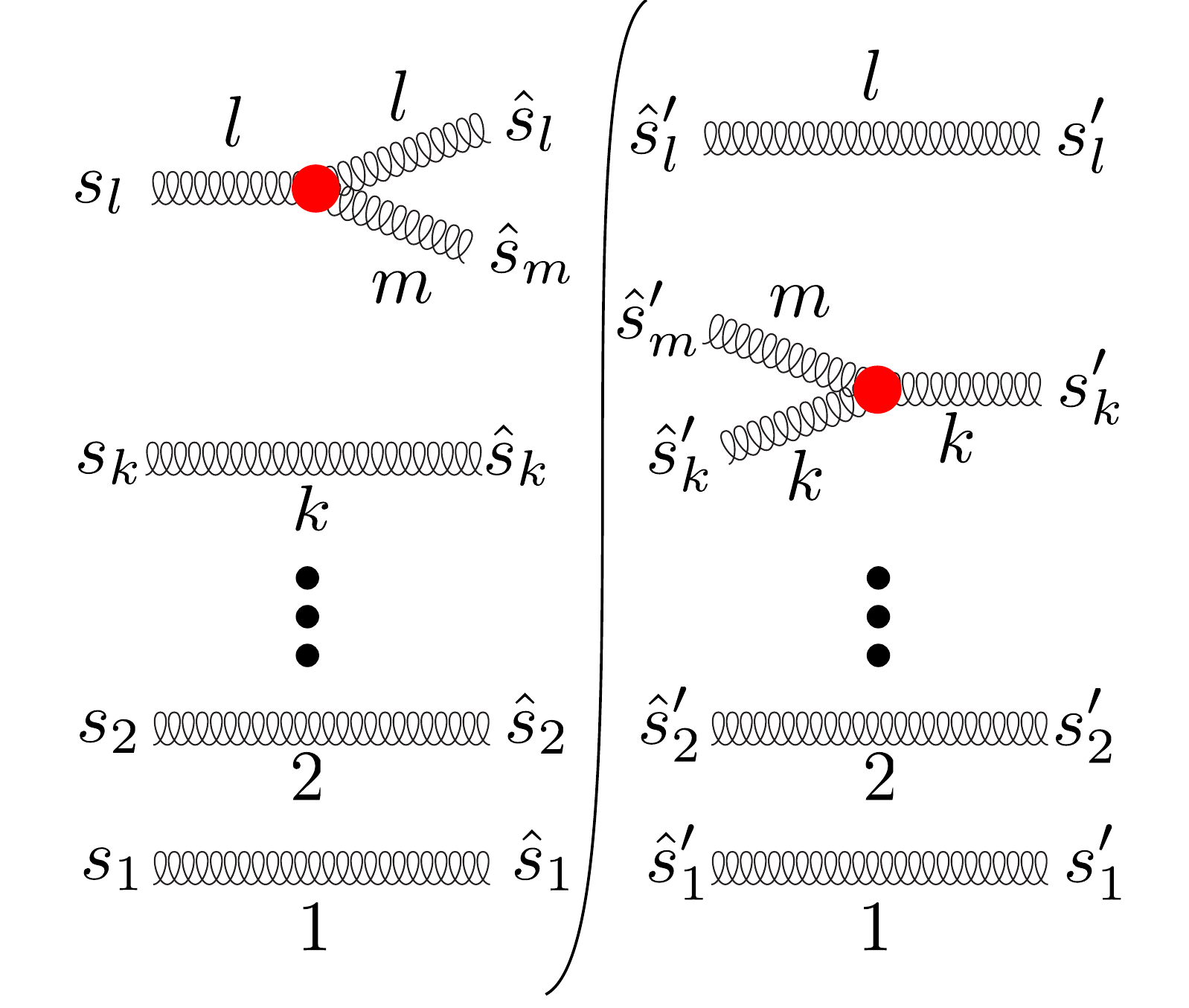}
\hskip 0.5 cm
\includegraphics[width = 7 cm]{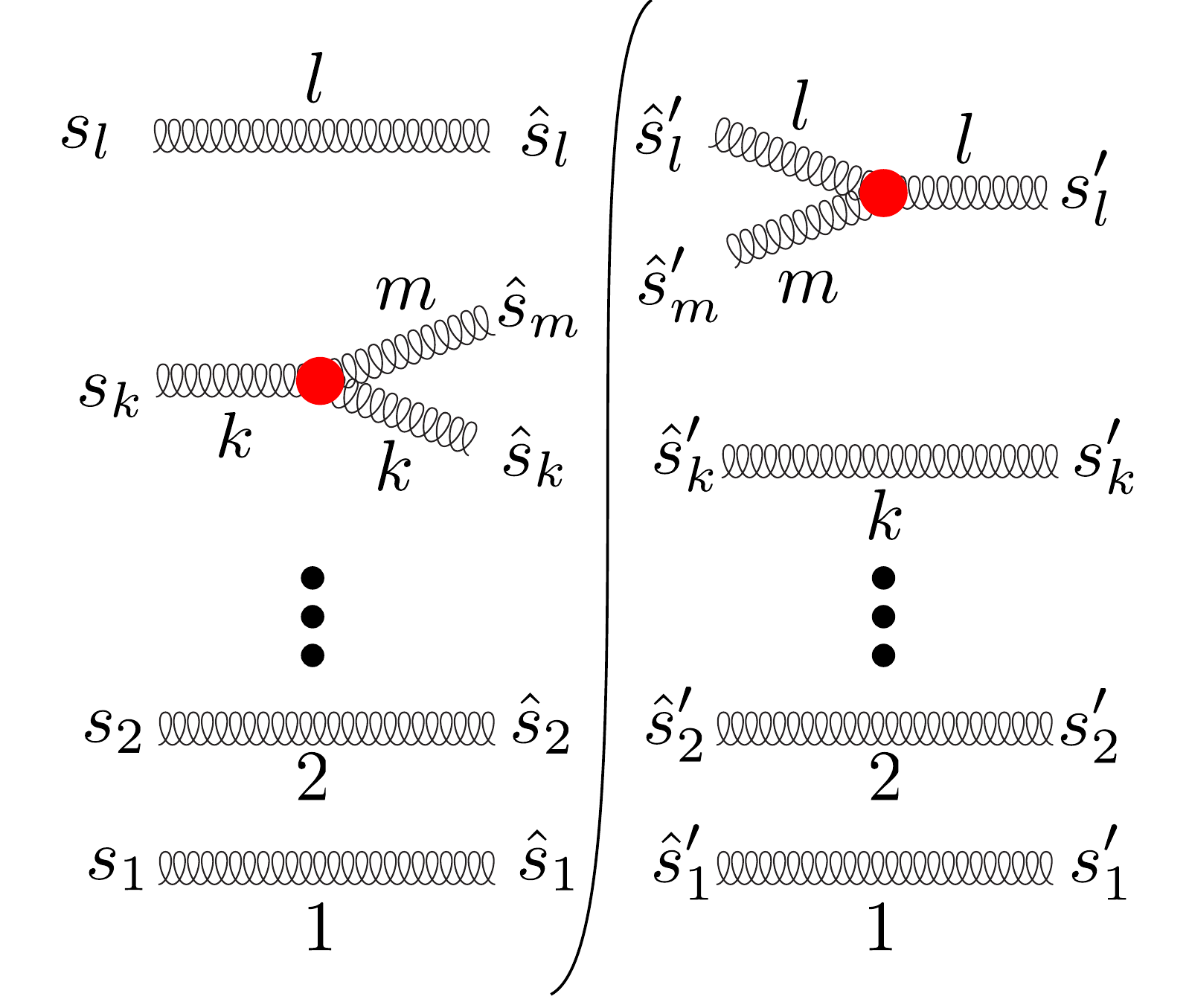}}
\caption{Illustration of the functions $W_{lk}$ and $W_{kl}$ in Eq.~(\ref{eq:Ylkdef}). The left hand diagram illustrates $W_{lk}$, in which parton $m$ is emitted from parton $l$ in the quantum amplitude and from parton $k$ in the conjugate amplitude. The right hand diagram illustrates $W_{kl}$, in which parton $m$ is emitted from parton $k$ in the quantum amplitude and from parton $l$ in the conjugate amplitude.}
\label{fig:Wlk}
}

In Eq.~(\ref{eq:Ylkdef}), the operator ${\cal A}_{lk}$ tells how the $l$-$k$ interference graphs are partitioned between a term treated as a splitting of parton $l$, with parton $k$ playing an auxiliary role, and a term treated as a splitting of parton $k$, with parton $l$ playing an auxiliary role. In this paper, we allow ${\cal A}_{lk}$ to depend on the spin indices of parton $m$, so that it is an operator on the parton spin space. In Ref.~\cite{NSspinless}, we specified $A_{lk}(\{\hat p\}_{m})$ as a spin independent function of the momenta. We give possibilities for ${\cal A}_{lk}$ in Sec.~\ref{sec:Alk}. The structure of the calculation outlined in this and the following section does not depend on which of these choices of ${\cal A}_{lk}$ one takes, although certainly the numerical performance of the algorithm can depend on the choice. We when we provide numerical examples in Secs.~\ref{sec:example1} and \ref{sec:3steps}, we make a definite choice that will be explained in Sec.~\ref{sec:Alk}.

In Eq.~(\ref{eq:HLeadingColor}), the spin matrix element of ${\cal Y}(l,k;\{\hat f,\hat p\}_{m})$ multiplies a spin independent function $\Phi_{lk}$. This function appears also in the denominator of ${\cal Y}(l,k;\{\hat f,\hat p\}_{m})$, so that it cancels. The function $\Phi_{lk}$ is the spin averaged splitting function, the spin average of the spin dependent splitting functions in ${\cal Y}(l,k;\{\hat f,\hat p\}_{m})$, defined by setting $\{\hat s'\}_m = \{\hat s\}_m$ and summing over these after-splitting spins and setting $\{s'\}_{m-1} = \{s\}_{m-1}$ and averaging over these before-splitting spins. The spin matrix elements of ${\cal Y}$ are the ratios of the splitting functions with spin to the spin-averaged splitting functions. Thus they are normalized to
\begin{equation}
\label{eq:Ynormalization}
\sum_{\{\hat s\}_m} \frac{1}{2^{m+1}}\sum_{\{s\}_{m-1}}
\sbra{\{\hat s,\hat s\}_{m}}
{\cal Y}(l,k;\{\hat f,\hat p\}_{m})
\sket{\{s,s\}_m}
= 1
\;\;.
\end{equation}
The functions $\Phi_{lk}$ are given in Ref.~\cite{NSspinless}. They can be recovered from the matrix elements of ${\cal Y}$ by using the normalization condition (\ref{eq:Ynormalization}).

The spin matrix element of ${\cal Y}(l,k;\{\hat f,\hat p\}_{m})$ multiplies a color factor $\bra{\{\hat c\}_{m}} a^\dagger_{lk}\ket{\{c\}_{m-1}}$. This factor equals 1 provided two conditions hold. First, partons $l$ and $k$ must be color connected in the initial color state $\{c\}_{m-1}$. Second, the new color state $\{\hat c\}_{m}$ must be the same as $\{c\}_{m-1}$ with the gluon with label $m$ inserted between partons $l$ and $k$. If either of these conditions fails, this factor vanishes.

This completes the brief description of the factors in the term containing ${\cal Y}(l,k;\{\hat f,\hat p\}_{m})$ in Eq.~(\ref{eq:HLeadingColor}). There is a second term, which covers the possibility that the newly created final state parton, labeled $m$, is not a gluon. Then there is no interference diagram to consider so there is no second parton with label $k$. Our notation for this case uses functions with the same names as for when there is a helper parton $k$, but sets $k \to l$. One can have $f_{m} \ne \mathrm{g}$ in two cases.  In the first case, one has an initial state splitting in which $\{f_l,\hat f_l,\hat f_m\}$ is either $\{q,{\rm g},q\}$ or $\{\bar q,{\rm g},\bar q\}$. The color factor for this splitting, $\bra{\{\hat c\}_{m}} a^\dagger_{\rm g}(l)\ket{\{c\}_{m-1}}$, is 1 if the new color state $\{\hat c\}_m$ is the same as $\{c\}_{m-1}$ with the end of the string that was at quark or antiquark $l$ now terminated at quark or antiquark $m$ and the new gluon with label $l$ inserted just next to the end of the string. Otherwise, this factor vanishes. In the second case, one has an initial or final state splitting in which $\{f_l,\hat f_l,\hat f_m\}$ is either $\{{\rm g},q,\bar q\}$ or $\{{\rm g},\bar q,q\}$. The color factor $\bra{\{\hat c\}_{m}} a^\dagger_{q}(l)\ket{\{c\}_{m-1}}$ is 1 if the color state $\{\hat c\}_{m}$ is related to $\{c\}_{m-1}$ by cutting the color string on which parton $l$ lies into two strings, terminating at the new quark and antiquark. Otherwise, this factor vanishes.

For $k = l$, the definition of ${\cal Y}(l,l;\{\hat f,\hat p\}_{m})$ is very simple,
\begin{equation}
\label{eq:Ylldef}
{\cal Y}(l,l;\{\hat f,\hat p\}_{m}) =
\frac{C(\hat f_l,\hat f_m)}{\Phi_{ll}(\{\hat p,\hat f\}_{m})}\
{\cal W}(l,l;\{\hat f,\hat p\}_{m})
\;\;.
\end{equation}
The operators ${\cal W}$ are specified in Ref.~\cite{NSshower}. The color factor $C(\hat f_l,\hat f_m)$ corresponds to the color operators in the two cases in Eq.~(\ref{eq:HLeadingColor}) with $\hat f_m \ne {\rm g}$
\begin{equation}
\label{eq:Cfldef}
C(\hat f_l,\hat f_m) = 
\begin{cases}
C_{\rm F} & \{\hat f_l,\hat f_m\} = \{{\rm g},\bar q\} \ {\rm or}\ \{{\rm g},q\} \\
T_{\rm R} & \{\hat f_l,\hat f_m\} = \{q,\bar q\} \ {\rm or}\ \{\bar q,q\}
\end{cases}
\;\;.
\end{equation}
Again, the functions $\Phi_{ll}$ are the spin averaged splitting functions. They are given in Ref.~\cite{NSspinless}. They can be recovered from the matrix elements of ${\cal Y}$ by using the normalization condition (\ref{eq:Ynormalization}) with $k = l$.

We can think of the operators $\cal Y$ as generating evolution in spin. They operate on the space of spin states with two spin indices for each parton, with basis states $\sket{\{s',s\}_{m}}$. These operators map the spin state for $m-1$ final state partons into a spin state for $m$ final state partons. In general, the spin states are linear combinations, $\sket{\rho_{\rm spin}}$, of the basis states,
\begin{equation}
\sket{\rho_{\rm spin}^{(m)}} =
\sum_{\{s',s\}_{m}} \sket{\{s',s\}_m}
\sbrax{\{s',s\}_m}\sket{\rho_{\rm spin}^{(m)}}
\;\;.
\end{equation}
We will use subscripts ``spin'' for vectors in the spin space other than the basis vectors $\sket{\{s',s\}_m}$ in order to emphasize that these are vectors in the finite dimensional space with $2\times 2$ dimensions for each parton.

\subsection{Starting point with spin}

The starting point for evolution is a state for two final state partons and two initial state partons, assuming that we start with a $2 \to 2$ hard process. Such a state is a mixture of the basis states $\sket{\{p,f,c,s',s\}_{2}}$ that represent a complete description of the momenta, flavors, colors and spins for the partons, including two spin indices but, since we work in the leading color approximation, only one color index. The leading color starting state is then $\sket{\rho^{\rm lc}(t)}$ at shower time $t=0$ and is represented as a linear combination of basis states as
\begin{equation}
\sket{\rho^{\rm lc}(0)} = \frac{1}{2!}
\int \big[d\{p,f,c\}_{2}\big]\
\sum_{\{s',s\}_2}
\sket{\{p,f,c,s',s\}_{2}}
\sbrax{\{p,f,c,s',s\}_{2}}\sket{\rho^{\rm lc}(0)}
\;\;.
\end{equation}
Here $\sbrax{\{p,f,c,s',s\}_{2}}\sket{\rho(0)}$ is obtained from the $2 \to 2$ matrix element,\footnote{As explained in Ref.~\cite{NSshower}, we should most properly project out the component of $\ket{\ME(\{p,f\}_{2})}$ that is proportional to a color basis state $\ket{\{c\}_{2}}$ by using a dual basis state $\dualL\bra{\{c\}_{2}}$, but in the leading color limit there is no distinction between the dual basis states and the ordinary basis states.}
\begin{equation}
\begin{split}
\label{eq:rhodef1}
\sbrax{\{p,f,c,s',s\}_{2}}\sket{\rho^{\rm lc}(0)} ={}&
\frac{f_{a/A}(\eta_{\La},\mu^{2}_{F})
f_{b/B}(\eta_{\Lb},\mu^{2}_{F})}
{4n_\Lc(a) n_\Lc(b)\,2\eta_{\La}\eta_{\Lb}p_\LA\!\cdot\!p_\LB}
\\&\times
\brax{\ME(\{p,f\}_{2})}\ket{\{s',c\}_{2}}
\brax{\{s,c\}_{2}}\ket{\ME(\{p,f\}_{2})}
\;\;.
\end{split}
\end{equation}
From Ref.~\cite{NSspinless}, we recall that the starting point for evolution averaged over spins (and in the leading color approximation) is
\begin{equation}
\begin{split}
\label{eq:rhodef0}
\sbrax{\{p,f,c\}_{2}}\sket{\rho^{(0)}(0)} ={}&
\frac{f_{a/A}(\eta_{\La},\mu^{2}_{F})
f_{b/B}(\eta_{\Lb},\mu^{2}_{F})}
{4n_\Lc(a) n_\Lc(b)\,2\eta_{\La}\eta_{\Lb}p_\LA\!\cdot\!p_\LB}\
\sum_{\{s\}_2}
\big|
\brax{\{s,c\}_{2}}\ket{\ME(\{p,f\}_{2})}
\big|^2
\;\;.
\end{split}
\end{equation}
It is convenient to define an initial vector in the spin space, $\sket{\rho(\{p,f,c\}_2)_{\rm spin}}$ that depends on the momenta, flavors, and colors of the partons by
\begin{equation}
\begin{split}
\label{eq:rho2}
\sbrax{\{s',s\}_2}\sket{\rho(\{p,f,c\}_2)_{\rm spin}} ={}&
\frac{\sbrax{\{p,f,c,s',s\}_{2}}\sket{\rho^{\rm lc}(0)}}
{\sbrax{\{p,f,c\}_{2}}\sket{\rho^{(0)}(0)}}
\\
={}& 
\frac{\brax{\ME(\{p,f\}_{2})}\ket{\{s',c\}_{2}}
\brax{\{s,c\}_{2}}\ket{\ME(\{p,f\}_{2})}}
{\sum_{\{s\}_2}
\big|\brax{\{s,c\}_{2}}\ket{\ME(\{p,f\}_{2})}\big|^2}
\;\;.
\end{split}
\end{equation}
With this notation, we can write the initial state with spin as a product of the initial state without spin and a factor that contains an initial vector in the spin space,
\begin{equation}
\sbrax{\{p,f,c,s',s\}_{2}}\sket{\rho^{\rm lc}(0)}
=
\sbrax{\{p,f,c\}_{2}}\sket{\rho^{(0)}(0)}
\times
\sbrax{\{s',s\}_2}\sket{\rho(\{p,f,c\}_2)_{\rm spin}}
\;\;.
\end{equation}

\subsection{Evolution with spin}

The evolution equation with spin is closely related to the evolution equation averaged over spin at each step. To obtain the spin averaged evolution equation from Eq.~(\ref{eq:evolutiondetaillc}), we need to eliminate the spin indices and the sum over spins. Then in Eq.~(\ref{eq:HLeadingColor}) we need to replace the matrix elements of the operators ${\cal Y}$ on the spin space by 1, leaving only the spin averaged splitting functions $\Phi$. We can get the full spin dependence back in each evolution step by inserting a factor
\begin{equation}
\label{eq:Yfactor}
\sum_{\{\hat s',\hat s\}_{m}} 
\sket{\{\hat s',\hat s\}_{m}}
\sbra{\{\hat s',\hat s\}_{m}}
{\cal Y}(l,k;\{\hat f,\hat p\}_{m})
\;\;.
\end{equation}
Here the momenta and flavors $\{\hat f,\hat p\}_{m}$ are the momenta and flavors that appear in the integrations for the spinless evolution. In an implementation of the integrations as a Markov chain, they are the momenta and flavors chosen at that step. Similarly, $l$ is the index designating the parton that split at that step. If the splitting involved the emission of a gluon into the final state, then a partner parton with index $k$ is also selected. If not, then in that step $k = l$. 

We have displayed the spin indices in Eq.~(\ref{eq:Yfactor}), but we can recognize that the completeness relation for our basis states allows us to replace
\begin{equation}
\label{eq:spinsums}
\sum_{\{\hat s',\hat s\}_{m}} 
\sket{\{\hat s',\hat s\}_{m}}
\sbra{\{\hat s',\hat s\}_{m}}
= 1
\;\;.
\end{equation}
Thus we obtain the full spin dependence by inserting the operator ${\cal Y}(l,k;\{\hat f,\hat p\}_{m})$ corresponding to the parameters of that splitting. After several steps, the spin state is
\begin{equation}
\sket{\rho^{(m)}_{\rm spin}}
\equiv
{\cal Y}(l_m,k_m;\{f,p\}_{m}^{(m)})\cdots
{\cal Y}(l_4,k_4;\{f,p\}_{4}^{(4)})\,
{\cal Y}(l_3,k_3;\{f,p\}_{3}^{(3)})
\sket{\rho(\{p,f,c\}_2)_{\rm spin}}
\;\;.
\end{equation}
Here $\sket{\rho(\{p,f,c\}_2)_{\rm spin}}$ is the starting spin state obtained from the hard matrix element. The first splitting produces 3 final state partons, with momenta and flavors $\{f,p\}_{3}^{(3)}$, by splitting parton $l_3$ with the participation of partner parton $k_3$. The number of final state partons increases as the shower progresses, so that at the later stage there are $m$ partons.

\subsection{End of the shower with spin}

At the end of the shower, at shower time $t_{\rm f}$, we apply an evolution operator that turns the partons into hadrons, then measure the hadronic final state with a measurement function $F_{\rm h}$. Following Ref.~\cite{NSshower}, we write this as
\begin{equation}
\begin{split}
\label{eq:hadronization}
\sigma^{\rm lc}[F_{\rm h}] ={}& 
\sum_N
\frac{1}{N!}
\int \big[d\{p,f,c\}_{N}\big]\
\sbra{F_{\rm h}}{\cal U}^{\rm had}(\infty,t_{\rm f})
\sket{\{p,f,c\}_{N}}
\\ & \times
\sum_{\{s\}_N}
\sbrax{\{p,f,c,s,s\}_{N}}
\sket{\rho^{\rm lc}(t_{\rm f})}
\;\;.
\end{split}
\end{equation}
Here we have made the assumption that neither the hadronization process nor the ultimate hadronic measurement depends on parton spins. Thus we have computed the total probability to get momenta, flavors, and colors $\{p,f,c\}_{N}$ by setting $s = s'$ and summing over $s$. 

At each step of the evolution,
\begin{equation}
\sbrax{\{p,f,c,s',s\}_{m}}
\sket{\rho^{\rm lc}(t_{m})}
=
\sbrax{\{p,f,c\}_{m}}
\sket{\rho^{\rm (0)}(t_{m})}\times
\sbrax{\{s',s\}_{m}}
\sket{\rho^{(m)}_{\rm spin}}
\;\;.
\end{equation}
Thus,
\begin{equation}
\begin{split}
\label{eq:finalwithspin}
\sigma^{\rm lc}[F_{\rm h}] ={}& 
\sum_N
\frac{1}{N!}
\int \big[d\{p,f,c\}_{N}\big]\
\sbra{F_{\rm h}}{\cal U}^{\rm had}(\infty,t_{\rm f})
\sket{\{p,f,c\}_{N}}\
\sbrax{\{p,f,c\}_{N}}
\sket{\rho^{(0)}(t_{\rm f})}
\\ & \times
\sbrax{1_{\rm spin}}
\sket{\rho^{(N)}_{\rm spin}}
\;\;,
\end{split}
\end{equation}
where
\begin{equation}
\sbrax{1_{\rm spin}}\sket{\{s',s\}_{m}} = \prod_{i\in \{\La,\Lb,1,\dots,m\}}
\delta_{s'_i,s_i}
\;\;,
\end{equation}
so that
\begin{equation}
\sbrax{1_{\rm spin}}
\sket{\rho^{(N)}_{\rm spin}}
= 
\sum_{\{s',s\}_N}
\sbrax{1_{\rm spin}}\sket{\{s',s\}_{N}}
\sbrax{\{s',s\}_{N}}
\sket{\rho^{(N)}_{\rm spin}}
=\sum_{\{s\}_N}
\sbrax{\{s,s\}_{N}}
\sket{\rho^{(N)}_{\rm spin}}
\;\;.%
\end{equation}

According to Eq.~(\ref{eq:finalwithspin}), the leading color cross section including the effects of spin on the distribution of partons is the same as the leading color cross section without spin times a factor $\sbrax{1_{\rm spin}}
\sket{\rho^{(N)}_{\rm spin}}$. We propose to include this factor as a weight for each Monte Carlo event. In the next section, we investigate how to compute it.

\section{Evaluating the spin factor}
\label{sec:spineval}

We need to evaluate the spin factor in Eq.~(\ref{eq:finalwithspin}),
\begin{equation}
\begin{split}
\label{eq:spinfactor}
\sbrax{1_{\rm spin}}
\sket{\rho^{(N)}_{\rm spin}}
={}&
\sbra{1_{\rm spin}}
{\cal Y}(l_N,k_N;\{f,p\}_{N}^{(N)})\cdots
\\&\times
{\cal Y}(l_4,k_4;\{f,p\}_{4}^{(4)})\,
{\cal Y}(l_3,k_3;\{f,p\}_{3}^{(3)})
\sket{\rho(\{p,f,c\}_2)_{\rm spin}}
\;\;.
\end{split}
\end{equation}
This factor contains matrix multiplications, which can be made manifest by inserting spin sums using Eq.~(\ref{eq:spinsums}). How can we evaluate this spin factor? A little reflection suggests that a Monte Carlo summation, choosing the spin indices $s_i$ at random, is not a promising approach. This observation suggests performing all of the sums exactly. However, $\sket{\rho^{(N)}_{\rm spin}}$ is a vector in a space with $2^{2(N+2)}$ dimensions. If the number of final state partons is, say, 100, this vector has more components than any available computer can hold. Thus one should be careful about how to arrange the calculation.

We will follow a method inspired by a work by J.~Collins \cite{JCCspin} on the subject of including spin in parton shower Monte Carlo programs. (See also Refs.~\cite{KnowlesSpin} and \cite{HerwigSpin}.) The problem considered in that paper did not include the interference graphs, which we want to include, and does not fit well with the way that we have organized a parton shower. However, we can use the main idea: that the matrix product in Eq.~(\ref{eq:spinfactor}) is simplest if we evaluate it from left to right.

To proceed, we denote
\begin{equation}
\label{eq:Yspindef}
\sbra{1_{\rm spin}}{\cal Y}(l_N,k_N;\{f,p\}_{N}^{(N)})\cdots
{\cal Y}(l_m,k_m;\{f,p\}_{m}^{(m)})
= \sbra{Y^{(m-1)}_{\rm spin}}
\;\;.
\end{equation}
Then our weight factor is
\begin{equation}
\sbrax{1_{\rm spin}}
\sket{\rho^{(N)}_{\rm spin}} =
\sbrax{Y^{(2)}_{\rm spin}}\sket{\rho(\{p,f,c\}_2)_{\rm spin}}
\;\;.
\end{equation}
Following Collins, we can call $\sbra{Y^{(m-1)}_{\rm spin}}$ the decay matrix in spin space. We can generate $\sbra{Y^{(2)}_{\rm spin}}$ recursively, using
\begin{equation}
\label{eq:recursion}
\sbra{Y^{(m-1)}_{\rm spin}} = 
\sbra{Y^{(m)}_{\rm spin}} 
{\cal Y}(l_m,k_m;\{f,p\}_{m}^{(m)})
\;\;.
\end{equation}
This seems horribly complicated, but it is not. We will see that $\sbra{Y^{(m)}_{\rm spin}}$ has the structure
\begin{equation}
\label{eq:spinstructure}
\sbrax{Y^{(m)}_{\rm spin}}\sket{\{s',s\}_m}
= \prod_{j} y_m^{(j)}(s'_j,s_j)
\;\;,
\end{equation}
as illustrated in Fig.~\ref{fig:YspinStructure}. That is, the complete decay matrix in spin space is a product of decay matrices for the individual partons. Even though $\sbra{Y^{(m)}_{\rm spin}}$ has $2^{2(m+2)}$ components, these components are determined by $m+2$ vectors with 4 components each.

\FIGURE{
\centerline{\includegraphics[width = 8 cm]{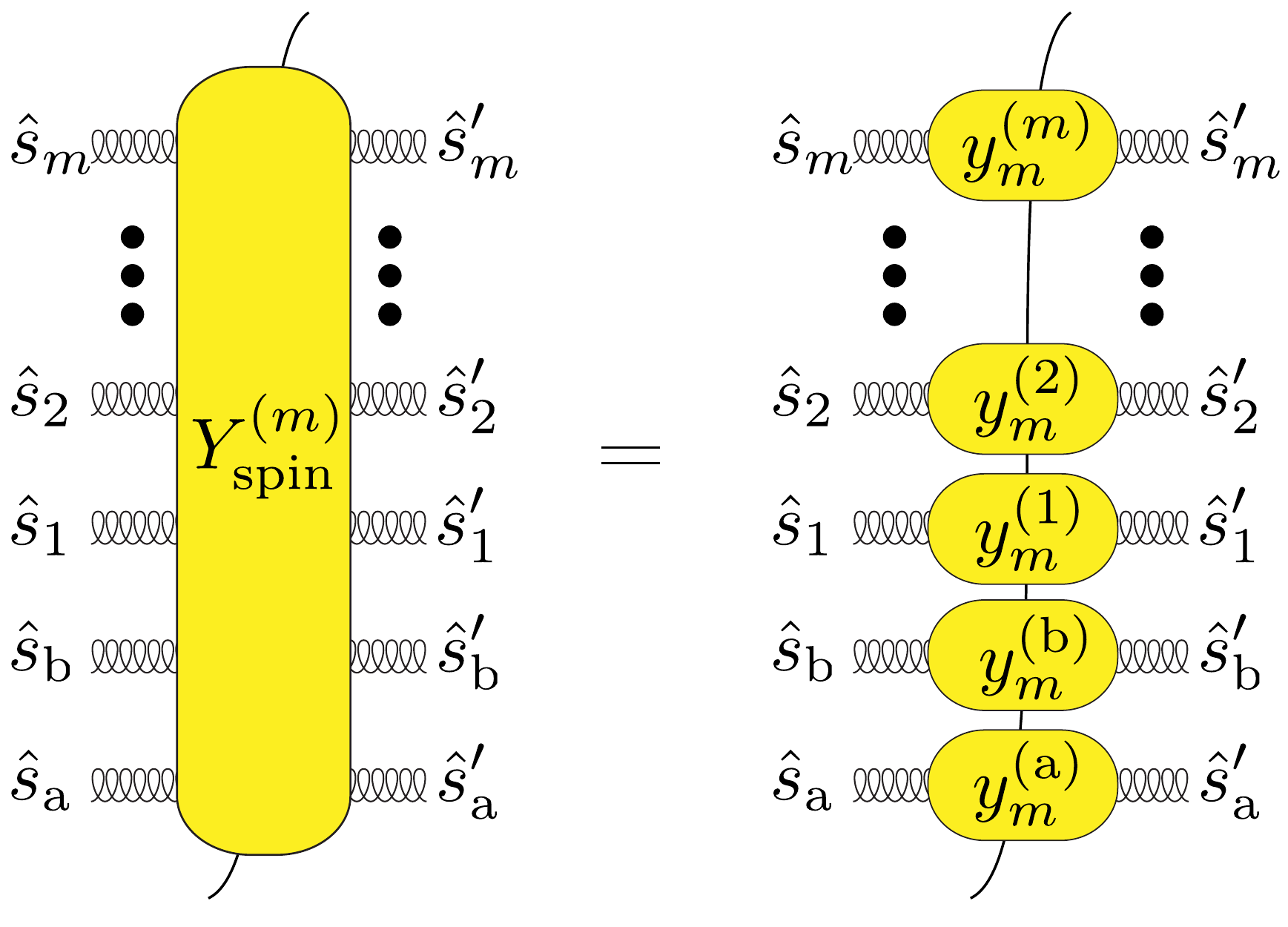}}
\caption{Structure of $\sbra{Y^{(N)}_{\rm spin}}$ as given in Eq.~(\ref{eq:spinstructure}).}
\label{fig:YspinStructure}
}

To see why Eq.~(\ref{eq:spinstructure}) holds, start with the initial condition for the recursion (\ref{eq:recursion}),
\begin{equation}
\sbrax{Y^{(N)}_{\rm spin}}\sket{\{s',s\}_{N}}
\equiv \sbrax{1_{\rm spin}}\sket{\{s',s\}_{N}}
=
\prod_{j}
\delta_{s'_j,s_j}
\;\;.
\end{equation}
This has the promised form\footnote{Eq.~(\ref{eq:startingspindecay}) is the simplest possibility, but one could use a hadronization model that results in a non-trivial spin decay matrix for each parton at the end of the shower. For instance, in a string model the polarization of a parton at the end of the shower could be correlated with the spatial directions of the string segments that couple to the parton.}  with
\begin{equation}
\label{eq:startingspindecay}
y_{N}^{(j)}(s'_j,s_j) = \delta_{s'_j,s_j}
\;\;.
\end{equation}

Next, we need to see whether this form is maintained under the recursion (\ref{eq:recursion}). For this we need the structure of the spin operators ${\cal Y}$ as given in Eqs.~(\ref{eq:Ylkdef}) and (\ref{eq:Ylldef}). This, in turn, requires us to examine the structure of the spin operators ${\cal W}$ as given in Ref.~\cite{NSshower} and illustrated in Figs.~\ref{fig:Wll} and \ref{fig:Wlk}. We first note that the spins of the partons other than $l$ and $k$, that is the partons not involved in the splitting at step $m$, are left unchanged under the splitting:
\begin{equation}
y_{m-1}^{(j)}(s'_j,s_j) = y_{m}^{(j)}(s'_j,s_j)\qquad j\notin \{l_m,k_m\}
\;\;.
\end{equation}
Next, we need to consider the spins of the partons that {\em are} involved in the splitting. 

Consider first the case that $k_m = l_m$, which arises when $\hat f_{m} \ne \mathrm{g}$. Then we use Eq.~(\ref{eq:Ylldef}) and use the spin operator ${\cal W}(l,l;\{\hat f,\hat p\}_{m})$ from Ref.~\cite{NSshower}. Writing simply $l$ for $l_m$, the result is
\begin{equation}
\begin{split}
\label{eq:newspinsimple}
y_{m-1}^{(l)}(s'_l,s_l)
={}&
\frac{C(\hat f_l,\hat f_m)\,S_l(\{\hat f\}_{m})}
{\Phi_{ll}(\{\hat p,\hat f\}_{m})}\
\sum_{\hat s'_{m},\hat s_{m}}
y_{m}^{(m)}(\hat s'_{m},\hat s_{m})\,
\sum_{\hat s'_l,\hat s_l}
y_{m}^{(l)}(\hat s'_l,\hat s_l)
\\ & \times
v_l(\{\hat p, \hat f\}_{m},\hat s_{m},\hat s_{l},s_l)\,
v_l^*(\{\hat p, \hat f\}_{m},\hat s'_{m},\hat s'_{l},s'_l)
\;\;.
\end{split}
\end{equation}
There is a color factor $C(\hat f_l,\hat f_m)$, Eq.~(\ref{eq:Cfldef}), and a statistical factor $S_l(\{\hat f\}_{m})$ equal to 1/2 for a final state ${\rm g} \to {\rm g}+{\rm g}$ splitting\footnote{The ${\rm g} \to {\rm g} + {\rm g}$ splitting is part Eq.~(\ref{eq:newspinglue}) below.} and 1 for any other allowed splitting, as given in Refs.~\cite{NSshower,NSspinless}. In the denominator, there is the spin-averaged splitting function $\Phi_{ll}$ described in Ref.~\cite{NSspinless}. The logic of this is simple. We have the amplitude for a parton of spin $s_l$ to split into partons with spins $\hat s_l$ and $\hat s_{m}$ times the complex conjugate amplitude for a parton of spin $s'_l$ to split into partons with spins $\hat s'_l$ and $\hat s'_{m}$. These multiply the spin decay matrices for the daughter partons $l$ and $m$. This equation is illustrated in Fig.~\ref{fig:spinevolve1}.

\FIGURE{
\centerline{\includegraphics[width = 10 cm]{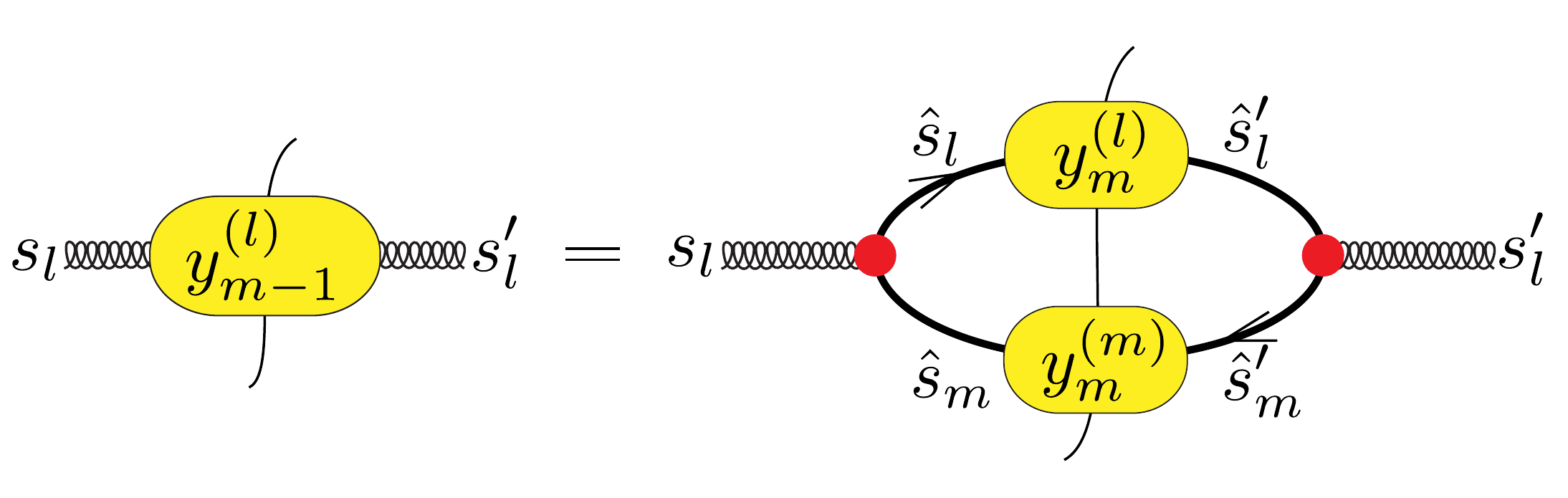}}
\caption{Illustration of Eq.~(\ref{eq:newspinsimple}) for spin evolution in a ${\rm g} \to q + \bar q$ splitting.}
\label{fig:spinevolve1}
}

The case that $k_m \ne l_m$, which arises when $\hat f_{m} = \mathrm{g}$, is a little more complicated. We use Eq.~(\ref{eq:Ylkdef}) and use the spin operators ${\cal W}(l,l;\{\hat f,\hat p\}_{m})$, ${\cal W}(l,k;\{\hat f,\hat p\}_{m})$, and ${\cal W}(k,l;\{\hat f,\hat p\}_{m})$ from Ref.~\cite{NSshower}. To keep the notation simple, we write $l$ for $l_m$ and $k$ for $k_m$. First, we note that the spin of the helper parton is not affected:
\begin{equation}
y_{m-1}^{(k)}(s'_k,s_k) = y_{m}^{(k)}(s'_k,s_k)
\;\;.
\end{equation}
Second, the new spin matrix for the parton, $l$, that splits is
\begin{equation}
\begin{split}
\label{eq:newspinglue}
y_{m-1}^{(l)}(s'_l,s_l)
={}&
\frac{C_{\rm F}\,S_l(\{\hat f\}_{m})}{\Phi_{lk}(\{\hat p,\hat f\}_{m})}
\sum_{\hat s'_{m},\hat s_{m}}
y_{m}^{(m)}(\hat s'_{m},\hat s_{m})\,
\sum_{\hat s'_l,\hat s_l}
y_{m}^{(l)}(\hat s'_l,\hat s_l)
\\ & \times
\bigg\{
v_{l}(\{\hat p, \hat f\}_{m},\hat s_{m},\hat s_{l},s_l)\,
v_{l}^*(\{\hat p, \hat f\}_{m},\hat s'_{m},\hat s'_{l},s'_l)
\\& + 
\theta(l\in \{1,\dots,m-1\}, \hat f_l = \hat f_{m} = {\rm g})
\\ & \times 
\left[
v_{2,l}(\{\hat p, \hat f\}_{m},\hat s_{m},\hat s_{l},s_l)\,
v_{2,l}^*(\{\hat p, \hat f\}_{m},\hat s'_{m},\hat s'_{l},s'_l)
\right.
\\ & \quad - 
\left.
v_{3,l}(\{\hat p, \hat f\}_{m},\hat s_{m},\hat s_{l},s_l)\,
v_{3,l}^*(\{\hat p, \hat f\}_{m},\hat s'_{m},\hat s'_{l},s'_l)
\right]
\\ & -
\frac{4\pi \alpha_{\rm s}\, A_{lk}(\{\hat p\}_{m},\hat s'_m,\hat s_m)}
{\hat p_{m}\!\cdot\!\hat p_l\ \hat p_{m}\!\cdot\!\hat p_k}\
\delta_{s_l \hat s_l}\, \delta_{s'_l \hat s'_l}
\\ & \times 
\left[
{\varepsilon(\hat p_{m},\hat s_{m};\hat Q)^* \!\cdot\!\hat p_l}\
{\varepsilon(\hat p_{m},\hat s'_{m};\hat Q) \!\cdot \!\hat p_k}
\right.
\\ & \quad+
\left.
{\varepsilon(\hat p_{m},\hat s_{m};\hat Q)^* \!\cdot\!\hat p_k}\
{\varepsilon(\hat p_{m},\hat s'_{m};\hat Q) \!\cdot \!\hat p_l}
\right]
\bigg\}
\;\;.
\end{split}
\end{equation}
As in Eq.~(\ref{eq:newspinsimple}), the new spin matrix for parton $l$ is constructed by a matrix multiplication from the old spin decay matrix of the daughter partons $l$ and $m$. There is a color factor $C_{\rm F}$ and a statistical factor $S_l(\{\hat f\}_{m})$. In the denominator, there is the spin-averaged splitting function $\Phi_{lk}$ described in Ref.~\cite{NSspinless}. All of this multiplies a factor in braces that is constructed from the splitting functions for the quantum amplitudes. The first term is the splitting amplitude for a parton of spin $s_l$ to split into partons with spins $\hat s_l$ and $\hat s_{m}$ times the complex conjugate amplitude for a parton of spin $s'_l$ to split into partons with spins $\hat s'_l$ and $\hat s'_{m}$. For a final state splitting in which both daughter partons are gluons, there is a correction term involving pieces $v_2$ and $v_3$ of the $\mathrm g \to \mathrm g + \mathrm g$ amplitude. Following Ref.~\cite{NSshower}, we write the ggg vertex as the sum of three terms,
\begin{equation}
\label{eq:vgg}
v^{\alpha \beta \gamma}(p_a, p_b, p_c)
= v_1^{\alpha \beta \gamma}(p_a, p_b, p_c)
+ v_2^{\alpha \beta \gamma}(p_a, p_b, p_c)
+ v_3^{\alpha \beta \gamma}(p_a, p_b, p_c)
\;\;,
\end{equation}
where
\begin{equation}
\begin{split}
\label{eq:vgg123}
v_1^{\alpha \beta \gamma}(p_a, p_b, p_c)
={}& g^{\alpha\beta} (p_a - p_b)^\gamma
\;\;,
\\
v_2^{\alpha \beta \gamma}(p_a, p_b, p_c)
={}& g^{\beta\gamma} (p_b - p_c)^\alpha
\;\;,
\\
v_3^{\alpha \beta \gamma}(p_a, p_b, p_c)
={}& g^{\gamma\alpha} (p_c - p_a)^\beta
\;\;.
\end{split}
\end{equation}
Then $v_{J,l}(\{\hat p, \hat f\}_{m},\hat s_{m},\hat s_{l},s_l)$ is the part of the splitting amplitude built from term $J$ in the ggg vertex, $v_J$. This construction from Ref.~\cite{NSshower} breaks the symmetry between gluons $l$ and $m$ and ensures that there is a singularity when daughter gluon $m$ is soft, but not when daughter gluon $l$ is soft. An alternative is to omit the additional terms and multiply both the spin dependent splitting function and the spin averaged splitting function by $\theta(z < 1/2)$. (Here $z$ is the momentum fraction of gluon $m$, defined, for instance, as in Ref.~\cite{NSspinless}.) The remainder of Eq.~(\ref{eq:newspinglue}) gives the contribution from the $l$-$k$ interference graphs. It is built from the amplitude for soft gluon emission in the eikonal approximation. In the eikonal approximation, the spin of parton $l$ remains undisturbed. The function $A_{lk}(\{\hat p\}_{m},s'_m,s_m)$ specifies how we partition the $l$-$k$ interference graphs into a fraction $A_{lk}$ associated with parton $l$ and a fraction $A_{kl}$ associated with parton $k$. In this paper, we allow $A_{kl}$ to depend on the spin indices $\hat s_m$ and $\hat s'_m$. One, spin independent, choice for this function was specified in Ref.~\cite{NSspinless}. We will give alternative definitions in Sec.~\ref{sec:Alk}. The general structure of the recursion relation Eq.~(\ref{eq:newspinglue}) is illustrated in Fig.~\ref{fig:spinevolve2}. This structure applies independently of the choice of $A_{lk}$.

\FIGURE{
\centerline{\includegraphics[width = 12 cm]{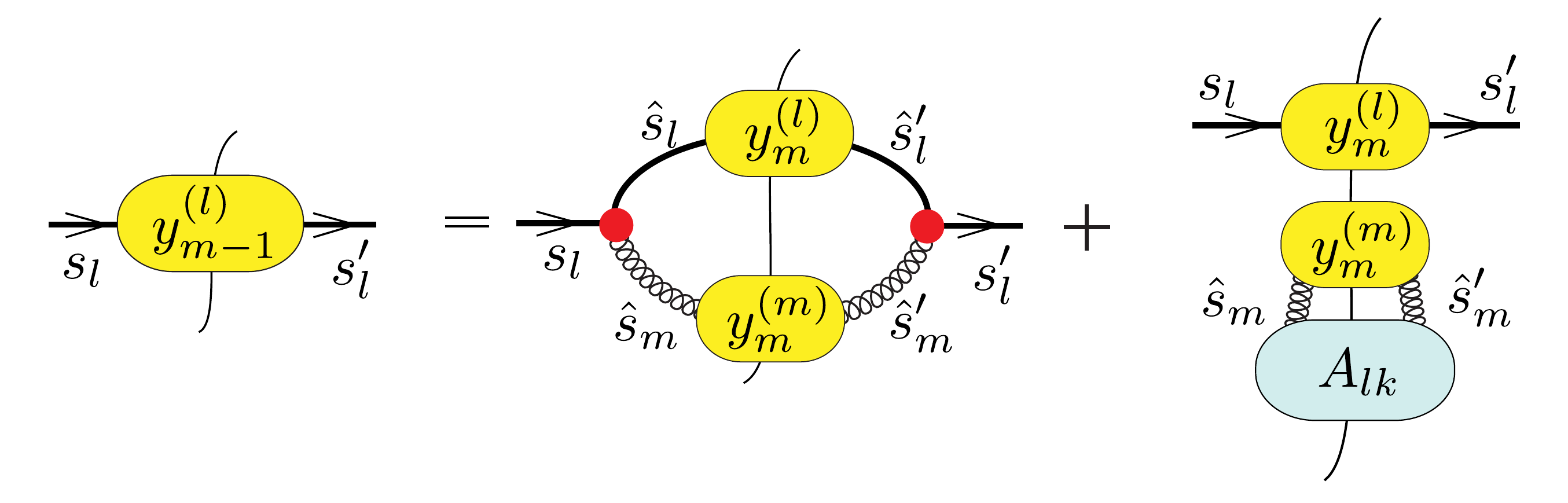}}
\caption{Illustration of Eq.~(\ref{eq:newspinglue}) for spin evolution in a $q \to q + {\rm g}$ splitting. The combination of the possibly spin dependent $A_{lk}$, the momenta $\hat p_l$ and $\hat p_k$, and the gluon polarization vectors for gluon $m$ are indicated in the drawing simply as an oval labeled $A_{lk}$.}
\label{fig:spinevolve2}
}

We see that the recursion relation (\ref{eq:recursion}) preserves the simple structure (\ref{eq:spinstructure}) that was present at the start of the recursion. Thus this structure holds at each step. We see further that at each step, $\sbra{Y^{(m-1)}_{\rm spin}}$ is obtained from 
$\sbra{Y^{(m)}_{\rm spin}}$ by a matrix multiplication involving a sum over four spin indices for each choice of $s'_l$ and $s_l$. At the end, we need a simple matrix multiplication to multiply $\sbra{Y^{(m)}_{\rm spin}}$ by $\sket{\rho(\{p,f,c\}_2)_{\rm spin}}$. From beginning to end, there are $N-1$ steps, where $N$ is the number of final state partons. Thus the amount of computer time and storage needed to execute the complete calculation is linear in the number of final state partons. One may encounter problems arising from the statistical fluctuations associated with the weight factors, but calculating the weight factors themselves should not be a problem.

\section{Properties of the spin decay matrices}
\label{sec:properties}

The spin decay matrices $y_{m}^{(l)}(s'_l,s_l)$ are hermitian,
\begin{equation}
y_{m}^{(l)}(s'_l,s_l)^* = y_{m}^{(l)}(s_l,s'_l)
\;\;.
\end{equation}
To see this, we simply note that the starting values of these matrices, $\delta_{s'_l,s_l}$, are hermitian and that the recursion relation specified by Eqs.~(\ref{eq:newspinglue}) and (\ref{eq:newspinglue}) preserves this property. (The coefficients $A_{lk}$ as constructed in Sec.~\ref{sec:Alk} obey $A_{lk}(\{p\}_m,\hat s'_m,\hat s_m)^* = A_{lk}(\{p\}_m,\hat s_m,\hat s'_m)$.) Thus the spin decay matrices have two eigenvectors and have real eigenvalues. The spin density matrix for the hard scattering, $\sbrax{\{s',s\}_2}\sket{\rho(\{p,f,c\}_2)_{\rm spin}}$ in Eq.~(\ref{eq:rho2}) is also hermitian, so the spin weight factor $\sbrax{1_{\rm spin}}
\sket{\rho^{(N)}_{\rm spin}} = \sbrax{Y^{(2)}_{\rm spin}}\sket{\rho(\{p,f,c\}_2)_{\rm spin}}$ is real.

If the recursion relation specified by Eqs.~(\ref{eq:newspinsimple}) and (\ref{eq:newspinglue}) were somewhat simpler, taking the form
\begin{equation}
\begin{split}
\label{eq:newspinsimplified}
y_{m-1}^{(l)}(s'_l,s_l)
={}&
C_l
\sum_{\hat s'_{m},\hat s_{m}}
y_{m}^{(m)}(\hat s'_{m},\hat s_{m})\,
\sum_{\hat s'_l,\hat s_l}
y_{m}^{(l)}(\hat s'_l,\hat s_l)
\\ & \times
v_l(\{\hat p, \hat f\}_{m},\hat s_{m},\hat s_{l},s_l)\,
v_l^*(\{\hat p, \hat f\}_{m},\hat s'_{m},\hat s'_{l},s'_l)
\;\;,
\end{split}
\end{equation}
where $C_l > 0$, then the spin decay matrices $y_{m}^{(l)}(s'_l,s_l)$ would be easily seen to be a positive matrix. That is both eigenvalues would be positive and $\sum u^*(s'_l)\, y_{m}^{(l)}(s'_l,s_l)\,u(s_l)$ would be positive for any vector $u$. To see this, we simply note that the starting values of these matrices, $\delta_{s'_l,s_l}$, are positive. Then we insert eigenvector expansions for the daughter parton spin decay matrices,
\begin{equation}
\begin{split}
y_{m}^{(l)}(\hat s'_{l},\hat s_{l}) ={}& 
\sum_{J}\lambda_{J,l}\,
\xi_{J,l}(\hat s'_{l})\,\xi_{J,l}^*(\hat s_{l})
\;\;,
\\
y_{m}^{(m)}(\hat s'_m,\hat s_m) ={}& 
\sum_{K}\lambda_{K,m}\,
\xi_{K,m}(\hat s'_{m})\,\xi_{K,m}^*(\hat s_{m})
\;\;,
\end{split}
\end{equation}
into Eq.~(\ref{eq:newspinsimplified}) and form a dot product with $u^*(s'_l)$ and $u(s_l)$, where $u$ is an arbitrary vector. This gives
\begin{equation}
\begin{split}
\label{eq:positivity}
\sum_{s'_l,s_l} u^*(s'_l)\, y_{m-1}^{(l)}(s'_l,s_l)\,u(s_l)
={}&
C_l
\sum_{J,K}\lambda_{J,l}\, \lambda_{K,m}
\\ & \times
\sum_{s_l,\hat s_m,\hat s_l}
u(s_l)\,v_l(\{\hat p, \hat f\}_{m},\hat s_{m},\hat s_{l},s_l)\,
\xi_{J,l}^*(\hat s_{l})\,\xi_{K,m}^*(\hat s_{m})
\\ & \times
\sum_{s'_l,\hat s'_m,\hat s'_l}
u^*(s'_l)\,v_l^*(\{\hat p, \hat f\}_{m},\hat s'_{m},\hat s'_{l},s'_l)\,
\xi_{J,l}(\hat s'_{l})\,\xi_{K,m}(\hat s'_{m})
\;\;.
\end{split}
\end{equation}
This is a sum of positive eigenvalues $\lambda_{J,l}$ and $\lambda_{K,m}$ times the square of the absolute value of a certain quantity. It is thus positive, showing that $y_{m-1}^{(l)}(s'_l,s_l)$ is a positive matrix. If the spin decay matrices are positive, then their product with the spin density matrix for the hard scattering, $\sbrax{\{s',s\}_2}\sket{\rho(\{p,f,c\}_2)_{\rm spin}}$ in Eq.~(\ref{eq:rho2}), must give a positive spin weight factor.

Since the main term in the recursion relation leads to positive spin decay matrices, one may reasonably suspect that the spin weight factor is positive for most events. However, since there are additional terms in the recursion relation (\ref{eq:newspinglue}), there can be configurations in which the spin weight factor has a negative eigenvalue.

\section{Example of one step in spin evolution}
\label{sec:example1}

We can gain some insight into the evolution of the spin decay matrices as we go backward from the end of the shower to the beginning. Let us write the spin decay matrix $y$ for the mother parton in the form
\begin{equation}
\label{eq:ydecomposition}
y_{m-1}^{(l)}(s'_l,s_l) = Y_l\left[
2P_l\, \xi_l(s'_l)\, \xi_l^*(s_l) 
+ (1-P_l)\,\delta_{s'_l,s_l}
\right]
\;\;.
\end{equation}
Here $\xi_l$ is a two component vector normalized to $\sum |\xi_l(s)|^2 = 1$. It is the eigenvector of the matrix $y$ with eigenvalue $\lambda_+ = Y_l\,(1+P_l)$. The other eigenvector, with eigenvalue $\lambda_- = Y_l\,(1-P_l)$, is orthogonal to $\xi$. We specify which eigenvector is $\xi$ by choosing $\lambda_+ \ge \lambda_-$, so that $P_l \ge 0$ (as long as $Y_l>0$). It will be helpful to have some names to apply to these variables. We can call $Y_l$ the spin enhancement factor and $P_l$ the fractional polarization of parton $l$. Then $\xi_l$ gives the direction of the polarization. This is a modification of the conventional language, in which ``fractional polarization'' and ``polarization vector'' refer to the preparation of a state rather than to its decay pattern, but it seems appropriate when we think of the spin structure as evolving backwards, toward the hard interaction.

For a gluon state, the polarization vector $\varepsilon$ corresponding to the vector $\xi$ is
\begin{equation}
\varepsilon^\mu = \sum_s \xi(s)\, \varepsilon^\mu(p,s;Q)
\;\;.
\end{equation}
Here the vectors $\varepsilon^\mu(p,s;Q)$ are helicity eigenvectors, orthogonal to $p$ and to the total final state momentum $Q$. We will be mostly interested in plane polarized states, for which
\begin{equation}
\xi(-1) = e^{i\phi}/\sqrt 2,\quad
\xi(+1) = e^{-i\phi}/\sqrt 2
\;\;.
\end{equation}
This makes $\varepsilon^\mu$ a real vector, orthogonal to $p$ and $Q$. It is rotated about $p$ (in the $\vec Q = 0$ frame) through angle $\phi$, starting from a plane that depends on the phase conventions used to define $\varepsilon^\mu(p,s;Q)$.

The mother parton spin decay matrix is determined by the spin decay matrices of the two daughter partons, which we parameterize as
\begin{equation}
\begin{split}
y_m^{(l)}(\hat s'_l,\hat s_l) ={}& \hat Y_l\left[
2\hat P_l\, \hat \xi_l(\hat s'_l)\, \hat \xi_l^*(\hat s_l) 
+ (1-\hat P_l)\,\delta_{\hat s'_l,\hat s_l}
\right]
\;\;,
\\
y_m^{(m)}(\hat s'_m,\hat s_m) ={}& \hat Y_m\left[
2\hat P_m\, \hat \xi_m(\hat s'_m)\, \hat \xi_m^*(\hat s_m) 
+ (1-\hat P_m)\,\delta_{\hat s'_m,\hat s_m}
\right]
\;\;.
\end{split}
\end{equation}

We consider the example of a final state ${\rm g} \to {\rm g} + {\rm g}$ splitting. The matrix $y_{m-1}^l(s_l,s'_l)$ depends on several parameters. For the sake of illustration, we make arbitrary choices for these with the aim of representing a roughly collinear splitting. With the convention that we denote momentum vectors by $p = (p^0,p^1,p^2,p^3)$, we take $Q = (1000,0,0,0)$. A gluon with momentum along the $x$ axis splits into two gluons with momenta
\begin{equation}
\begin{split}
\hat p_l ={}& (110.5,110,+10 \sin\phi,-10 \cos{\phi})
\;\;,
\\
\hat p_m ={}& (\ 90.6,\ \ 90,-10 \sin\phi,+10 \cos{\phi})
\;\;.
\end{split}
\end{equation}
Here $\phi$ gives the azimuthal angle of the splitting. We will let $\phi$ vary.
The splitting includes quantum interference in which gluon $m$ can also be emitted by a gluon $k$ with momentum
\begin{equation}
\hat p_k = (141.4,140,0,20)
\;\;.
\end{equation}
In this example, the interference contribution is not particularly large because gluon $m$ is not soft. We will let the two daughter partons carry 15\% linear polarization, in the direction $\hat \phi_l = 0.075\,\pi$ and $\hat \phi_m = - 0.075\,\pi$. Thus $\hat P_l = \hat P_m = 0.15$. We set $\hat Y_l = \hat Y_m = 1$.

We need a definite choice for the function $A_{lk}$ that partitions the interference graphs between a term associated with the splitting of parton $l$ and a term associated with the splitting of parton $k$. We adopt the definitions in Eqs.~(\ref{eq:Alkspindependent}) and (\ref{eq:Alkprimedef2}) from Sec.~\ref{sec:Alk}.

In Fig.~\ref{fig:Yfor15pol}, we plot the average spin enhancement factor $Y_l$ for this splitting as a function of the azimuthal angle $\phi$ of the splitting. We see that the enhancement factor varies about 15\% from 1, where 1 corresponds to no enhancement. The mother gluon acquires a net polarization in the splitting. The polarization $P_l$ is shown as a function of $\phi$ in Fig.~\ref{fig:polfor15pol}. The mother polarization depends on $\phi$ but its average is about 15\%, the input polarization of the daughter partons.

\FIGURE{
\centerline{\includegraphics[width = 6 cm]{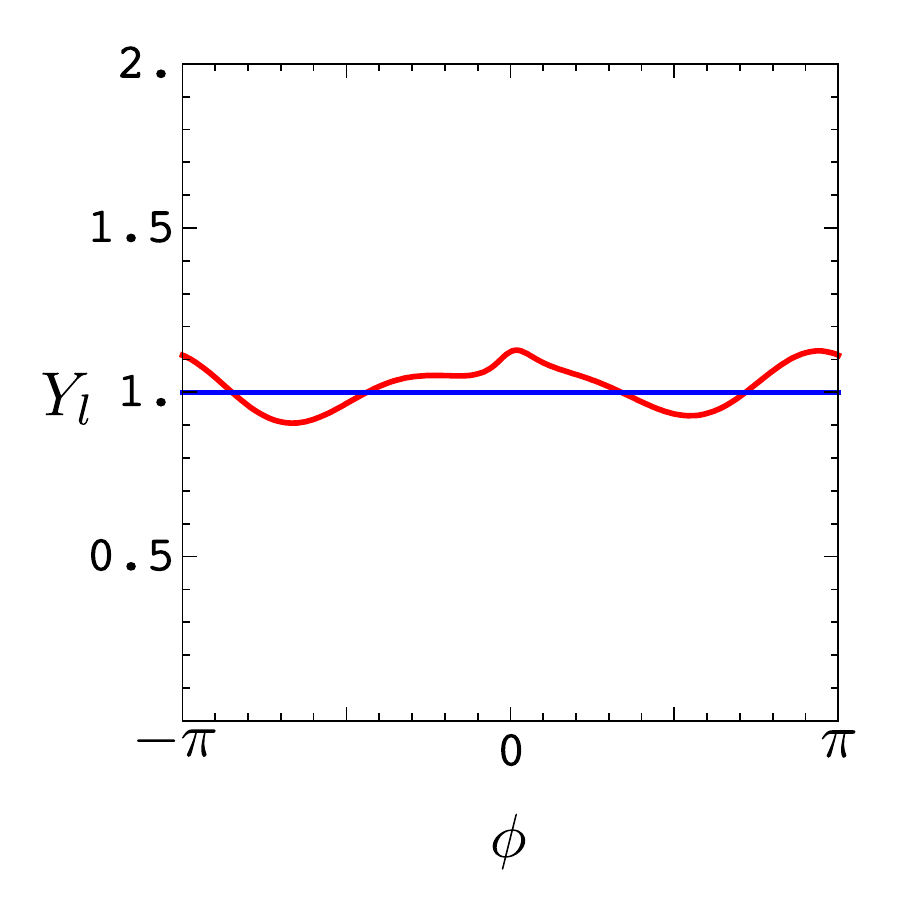}}
\caption{The spin enhancement factor $Y_l$, Eq.~(\ref{eq:ydecomposition}), contributed by a single splitting in which the daughter polarizations are 15\%. The splitting is roughly collinear, as described in the text. The spin enhancement factor is plotted against the azimuthal angle $\phi$ of the splitting along with a line indicating $Y_l=1$.}
\label{fig:Yfor15pol}
}

\FIGURE{
\centerline{\includegraphics[width = 6 cm]{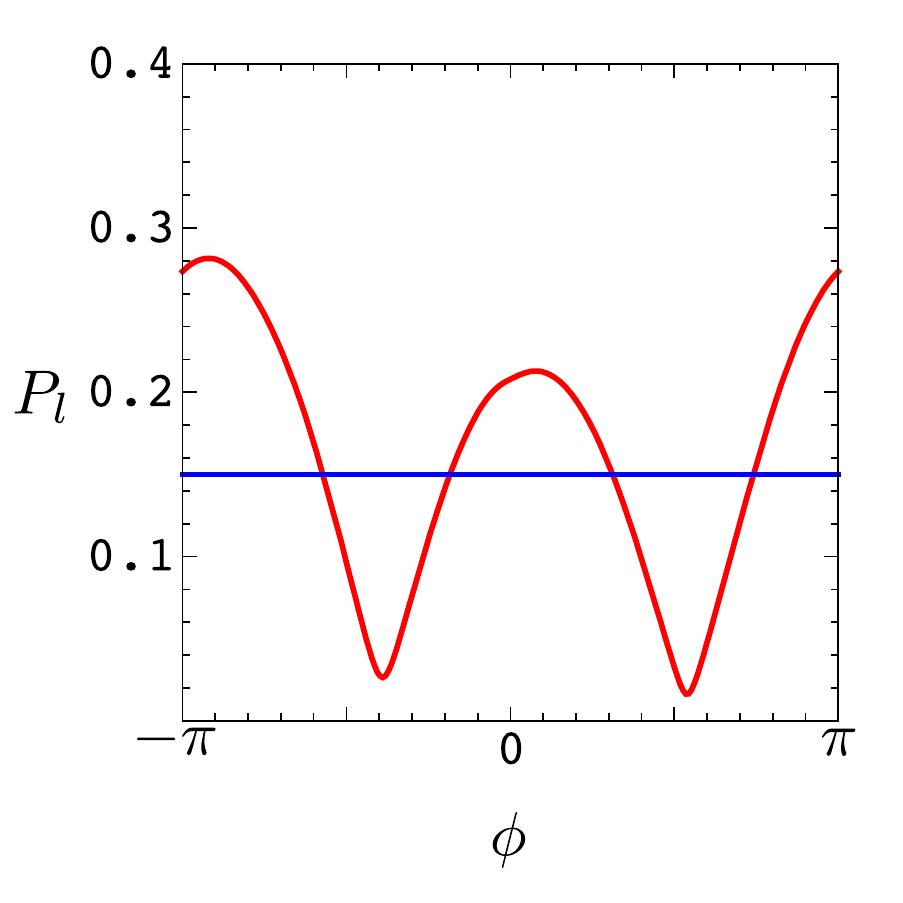}}
\caption{The mother parton polarization $P_l$, Eq.~(\ref{eq:ydecomposition}), for the splitting described in the text, plotted against the azimuthal angle $\phi$ of the splitting. We also show a line indicating $P_l = 0.15$, the input polarization of the daughters.}
\label{fig:polfor15pol}
}

In this example, we used daughter polarizations of 15\%. If the daughter polarizations are zero, then the enhancement factor, $Y_l$, is 1 for all azimuthal angles $\phi$. How, then, do the gluons get polarized? To investigate this, we set the daughter polarizations to 0\% and plot the resulting polarization $P_l$ of the mother versus the azimuthal angle $\phi$. The result is shown in Fig.~\ref{fig:polfor0pol}. We see that the mother polarization varies with $\phi$ and is generally a little larger than 10\%. The corresponding polarization vector is approximately in the direction of $k_\perp$, the part of $p_l - p_m$ orthogonal to $p_l + p_m$ and $Q$. 

We conclude that at the first step back from the final state, the gluons will become polarized. From there on, they will continue to be polarized and their polarizations will generate spin enhancements in the distributions of the azimuthal angles of the splittings beyond what is present in the spin averaged case. Our numerical experiments, like that presented here, suggest that the correlations in azimuthal angle between two successive splittings are of rather modest size. We expect, however, that correlations among several splittings, involving products of the spin factors $Y_l$, can be quite different from 1.

\FIGURE{
\centerline{\includegraphics[width = 6 cm]{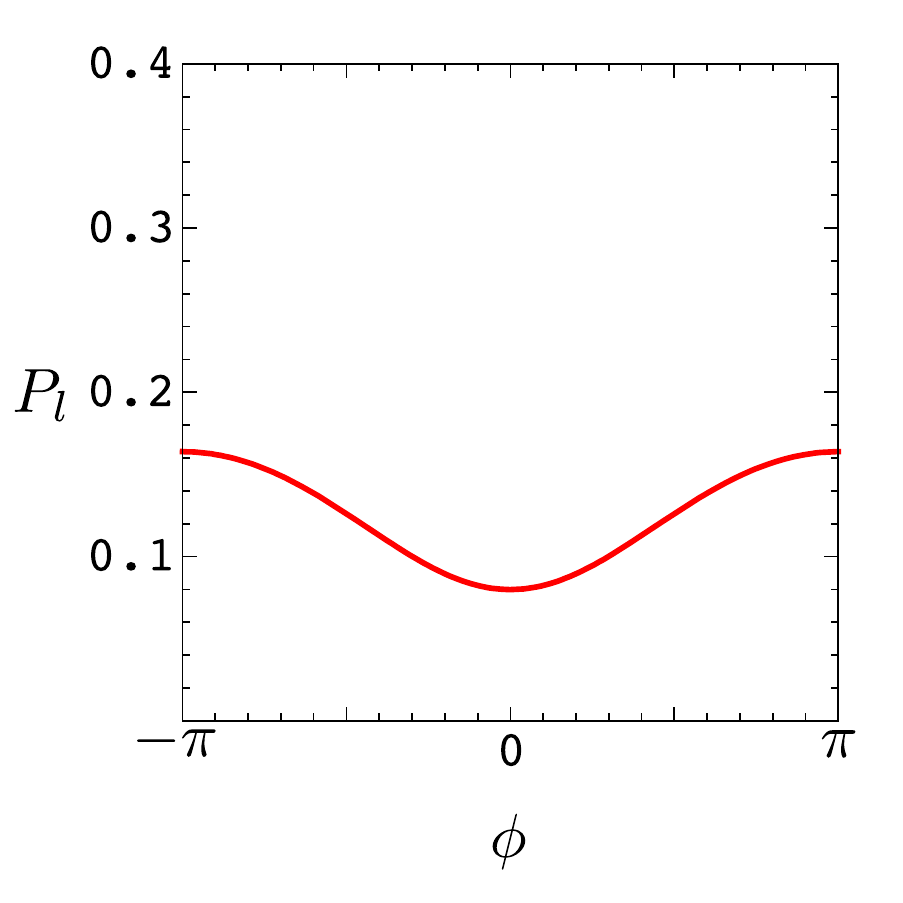}}
\caption{The mother parton polarization $P_l$, Eq.~(\ref{eq:ydecomposition}), for the splitting described in the text but with daughter polarizations set to zero. We plot $P_l$ against the azimuthal angle $\phi$ of the splitting. }
\label{fig:polfor0pol}
}

\section{Example of three steps in spin evolution}
\label{sec:3steps}

We can illustrate the calculation of the spin weight factor for a simple final state shower with three splittings. Suppose that shower evolution with spin averaged has produced a shower as illustrated in Fig.~\ref{fig:minishower}. The gluon with label 0 has split into gluons with labels 1 and 2 with the participation of gluon 7 in the soft interference term. Then gluon 1 has split into gluons 3 and 4 with the participation of gluon 2 in the interference term. Finally gluon 2 has split into gluons 5 and 6 with the participation of gluon 4 in the interference term. There are other final state partons that do not participate and are not shown. At each splitting, a small amount of momentum is taken from the partons that did not split according to the momentum mapping of Ref.~\cite{NSshower}.

Suppose that at the end of the shower the momenta of the final partons, along with the momentum of all the final state partons, $q$, are given in components $(E,p_x,p_y,p_z)$ by
\begin{equation}
\begin{split}
\hat p_3 ={}& (114.564, 110, 32, 1) 
\;\;,
\\
\hat p_4 ={}& (131.852, 130, 22, -1)
\;\;,
\\
\hat p_5 ={}& ( 90.05, 90, 3 \cos\phi, 3 \sin\phi)
\;\;,
\\
\hat p_6 ={}& (70.0643, 70 , -3  \cos\phi, -3 \sin\phi)
\;\;,
\\
\hat p_7 ={}& (131.909, 130, -22, -4)
\;\;,
\\
\hat q ={}& (1000, 0, 0, 0)
\;\;.
\end{split}
\end{equation}
Note that the splittings in this example are approximately collinear and that the azimuthal angle of the 5-6 splitting is denoted as $\phi$ and left variable.

We assume that the spins of the partons at the end of the shower are not measured, so that their spin decay matrices are unit matrices. We then compute the spin decay matrices of partons 1 and 2. Finally, we use these spin decay matrices to compute\footnote{We again define the function $A_{lk}$ that partitions the interference graphs using Eqs.~(\ref{eq:Alkspindependent}) and (\ref{eq:Alkprimedef2}).} the spin decay matrix $y^{(0)}$ of parton 0. From this we can use Eq.~(\ref{eq:ydecomposition}) to compute the polarization of parton 0 and its spin enhancement factor $Y_0$. If parton 0 came directly from the hard interaction, then we would take the trace of $y^{(0)}$ with the spin dependent hard matrix element to calculate the spin weight factor. Thus the complete spin weight factor contains the spin enhancement factor $Y_0$ of parton 0 as a factor. 

In Fig.~\ref{fig:Y0}, we plot $Y_0$ as a function of the azimuthal angle $\phi$ of the 5-6 splitting. We see that angles near 0 and $\pi$ get higher weights. This illustrates that spin effects act to correlate azimuthal angles of splittings even though the spins of the partons at the end of the shower are not measured. We also see that the effect with just four active final state partons is not large, a little bigger than 10\%. We should note, however, that the effects can be larger when the depth of the shower is bigger and more final state angles are correlated.

\FIGURE{
\centerline{\includegraphics[width = 4.5 cm]{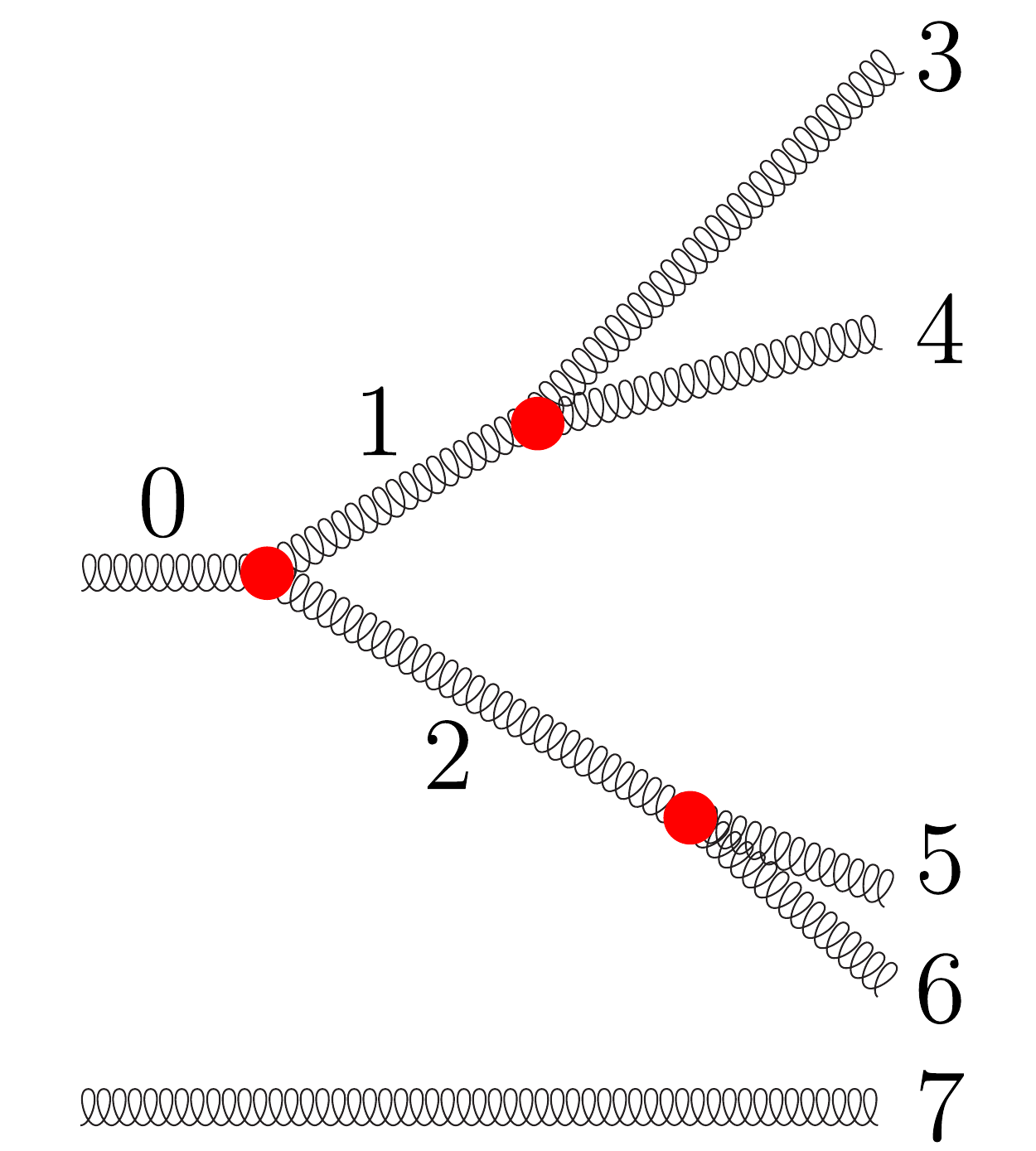}}
\caption{Example of a shower with three splittings, for which we calculate the spin enhancement factor $Y_0$ after working back to the first step.}
\label{fig:minishower}
}

\FIGURE{
\centerline{\includegraphics[width = 6 cm]{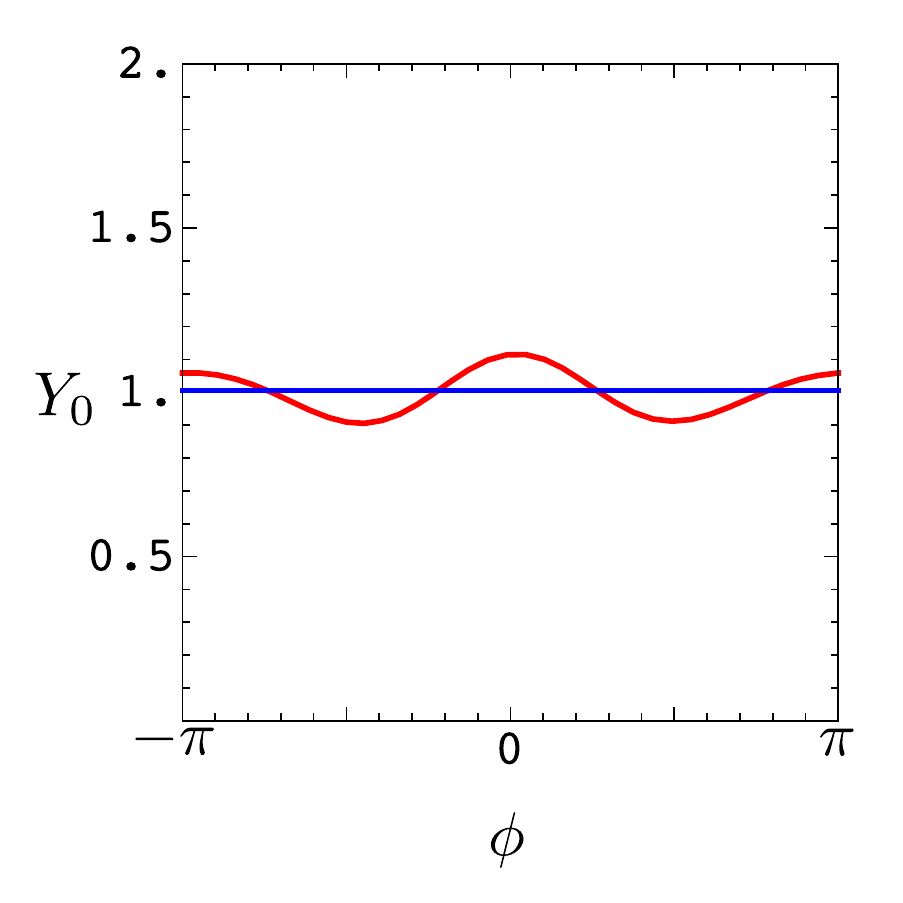}}
\caption{The spin enhancement factor $Y_0$ for the starting gluon in the mini-shower shown in Fig.~\ref{fig:minishower}. The spin enhancement factor is plotted against the azimuthal angle $\phi$ of the 5-6 splitting along with a line indicating $Y_0=1$.}
\label{fig:Y0}
}

\section{The dipole partitioning function}
\label{sec:Alk}

Eq.~(\ref{eq:newspinglue}) contains a function $A_{lk}(\{p\}_m,s'_m,s_m)$ that specifies how the two $l$-$k$ interference diagrams are partitioned into separate terms. A fraction  $A_{lk}$ is associated with the splitting of parton $l$ and comes with the momentum mapping for that splitting, while a fraction  $A_{kl}$ is associated with the splitting of parton $k$ and comes with the momentum mapping for that splitting. We have
\begin{equation}
\label{eq:Alksum}
A_{lk}(\{p\}_m,s'_m,s_m) + A_{kl}(\{p\}_m,s'_m,s_m) = 1
\;\;.
\end{equation}
We can generalize the choice of $A_{lk}(\{p\}_m)$ given in Ref.~\cite{NSspinless} in two ways. 

The first choice of $A_{lk}$ is spin independent,
\begin{equation}
A_{lk}(\{\hat p\}_{m}) 
= 
\frac{\hat p_{m}\!\cdot\!\hat p_l\ \hat p_{m}\!\cdot\!\hat p_k}
{2 \hat p_l\cdot D(\hat p_m,\hat Q)\cdot \hat p_k}
\left(
\frac{\hat p_l\cdot D(\hat p_m,\hat Q)\cdot \hat p_l}
{(\hat p_{m}\!\cdot\!\hat p_l)^{2}}
-A_{lk}'(\{\hat p\}_{m})\,
\frac{\hat P_{lk}\cdot D(\hat p_m,\hat Q)\cdot \hat P_{lk}}
{(\hat p_{m}\!\cdot\!\hat p_l\ \hat p_{m}\!\cdot\!\hat p_k)^{2}}
\right)\;\;,
\end{equation}
where $\hat Q$ is the total momentum of the final state partons, $\hat P_{lk}$ is the vector
\begin{equation}
\hat P_{lk} ={} \hat p_m\cdot \hat p_l\ \hat p_k
- \hat p_m \cdot \hat p_k\ \hat p_l
\;\;,
\end{equation}
$D(\hat p_m,\hat Q)^{\mu\nu}$ is the spin sum
\begin{equation}
\begin{split}
D(\hat p_m,\hat Q)^{\mu\nu} ={}& \sum_s 
\varepsilon^\mu(p_m,s;\hat Q)^*\,
\varepsilon^\nu(p_m,s;\hat Q)
\\
={}& - g^{\mu\nu} + \frac{\hat p_m^\mu \hat Q^\nu + \hat Q^\mu \hat p_m^\nu}
{\hat p_m\cdot \hat Q}
- \frac{\hat Q^2 \hat p_m^\mu \hat p_m^\nu}
{(\hat p_m\cdot \hat Q)^2}
\;\;,
\end{split}
\end{equation}
and $A'_{lk}$ is any positive function with
\begin{equation}
A'_{lk}(\{\hat p\}_{m}) + A'_{kl}(\{\hat p\}_{m}) =1
\;\;.
\end{equation}
Simple algebra starting from these relations yields the property (\ref{eq:Alksum}) of $A_{lk}$.

The choice of $A_{lk}$ affects the spin-averaged splitting function that appears in Eq.~(\ref{eq:HLeadingColor}),
\begin{equation}
\Phi_{lk}(\{\hat p,\hat f\}_{m}) = C_F[\overline W_{ll} - \overline W_{lk}]
\;\;,
\end{equation}
where $\overline W_{ll}$ and $\overline W_{lk}$ are the splitting functions averaged over spin, analyzed in Ref.~\cite{NSspinless}. As in Ref.~\cite{NSspinless}, it is useful to add and subtract the soft gluon approximation to $\overline W_{ll}$,
\begin{equation}
\overline W_{ll}^{\rm eikonal}
= 4 \pi \as\ \frac{\hat p_l \cdot D(\hat p_m, \hat Q)\cdot \hat p_l}
{(\hat p_m\cdot \hat p_l)^2}
\;\;.
\end{equation}
The $l$-$k$ interference function is already constructed using the eikonal approximation that gives the soft gluon limit and, with our definition, includes the function $A_{lk}$,
\begin{equation}
\overline W_{lk}
= 4 \pi \as\ 
A_{lk}\ \frac{\hat p_l \cdot D(\hat p_m, \hat Q)\cdot \hat p_k}
{\hat p_m\cdot \hat p_l\ \hat p_m\cdot \hat p_k}
\;\;.
\end{equation}
Thus we decompose $\Phi_{lk}$ into
\begin{equation}
\label{eq:Philk2}
\Phi_{lk}=
C_F\big[(\overline W_{ll} - \overline W_{ll}^{\rm eikonal})
+ (\overline W_{ll}^{\rm eikonal} - \overline W_{lk})\big]
\;\;.
\end{equation}
The term $(\overline W_{ll}^{\rm eikonal} - \overline W_{lk})$ includes the soft singularity and the soft$\times$collinear singularity, while $(\overline W_{ll} - \overline W_{ll}^{\rm eikonal})$ has only a collinear singularity. After a little bit of algebra, one obtains
\begin{equation}
\label{eq:Alkresult}
\overline W_{ll}^{\rm eikonal} - \overline W_{lk}
=
4 \pi \as\ 
A'_{lk}\
\frac{\hat P_{lk} \cdot D(\hat p_m, \hat Q)\cdot \hat P_{lk}}
{(\hat p_m\cdot \hat p_l\ \hat p_m\cdot \hat p_k)^2}
=
4 \pi \as\ 
A'_{lk}\
\frac{-\hat P_{lk}^2 }
{(\hat p_m\cdot \hat p_l\ \hat p_m\cdot \hat p_k)^2}
\;\;.
\end{equation}
The last equality here follows from the fact that $\hat p_m\cdot \hat P_{lk} = 0$. This has the feature that $(\overline W_{ll}^{\rm eikonal} - \overline W_{lk})$ is positive as long as $A'_{lk}$ is positive. This is important for constructing the spin-averaged shower as a Markov process.

The choice of $A'_{lk}$ in Ref.~\cite{NSspinless} was
\begin{equation}
\label{eq:Alkprimedef1}
A'_{lk}(\{\hat p\}_{m})
= \frac{
(p_m\cdot p_k)^2\
\hat p_l \cdot D(\hat p_m, \hat Q)\cdot \hat p_l}
{
(p_m\cdot p_k)^2\
\hat p_l \cdot D(\hat p_m, \hat Q)\cdot \hat p_l
+
(p_m\cdot p_l)^2\
\hat p_k \cdot D(\hat p_m, \hat Q)\cdot \hat p_k
}
\;\;.
\end{equation}
With this choice, $A_{lk} = A'_{lk}$. This choice has three good features. First, it vanishes $\hat p_m\cdot\hat p_k \to 0$ in such a way that there is no singularity in $(\overline W_{ll}^{\rm eikonal} - \overline W_{lk})$ when $\hat p_m$ becomes collinear to $\hat p_k$.\footnote{The singularity when $\hat p_m$ becomes collinear to $\hat p_l$ is assigned to $(\overline W_{kk}^{\rm eikonal} - \overline W_{kl})$.} Second, it is invariant under rescaling of $\hat p_m$. Third, it is also invariant under rescaling of $\hat p_l$ and of $\hat p_k$ so that, for the case of massless partons $l$ and $k$, it is a function only of the angles of the partons.

Another possible choice is
\begin{equation}
\label{eq:Alkprimedef2}
A'_{lk}(\{\hat p\}_{m})
= \frac{\hat p_{m}\cdot\hat p_k\ \hat p_{l}\cdot\hat Q}
{\hat p_{m}\cdot\hat p_k\ \hat p_{l}\cdot\hat Q
 + \hat p_{m}\cdot\hat p_l\ \hat p_{k}\cdot\hat Q}\;\;.
\end{equation}
This choice has the same three good properties, but it is simpler. A third possible choice is
\begin{equation}
\label{eq:Alkprimedef3}
A'_{lk}(\{\hat p\}_{m})
= \frac{\hat p_{m}\cdot\hat p_k}
{\hat p_{m}\cdot\hat p_k + \hat p_{m}\cdot\hat p_l}\;\;.
\end{equation}
This has two of the good properties but is not invariant under rescaling of $\hat p_l$ and $\hat p_k$, so that the partitioning depends on the energies as well as the angles of partons $l$ and $k$. This choice does have the advantage of being the simplest. It is analogous to the partitioning factor in the Catani-Seymour scheme for dipole subtractions \cite{CataniSeymour}.

One can also choose a spin dependent partitioning factor. This is particularly useful in the case, investigated in this paper, that we take the leading color approximation. We rewrite Eq.~(\ref{eq:newspinglue}) by adding and subtracting the eikonal approximation for the direct $l$-$l$ splitting graph and by using a spin dependent $A_{lk}$ as
\begin{equation}
\begin{split}
\label{eq:newspinglue2}
y_{m-1}^{(l)}(s'_l,s_l)
={}&
\frac{C_{\rm F}\,S_l(\{\hat f\}_{m})}{\Phi_{lk}(\{\hat p,\hat f\}_{m})}
\sum_{\hat s'_{m},\hat s_{m}}
y_{m}^{(m)}(\hat s'_{m},\hat s_{m})\,
\sum_{\hat s'_l,\hat s_l}
y_{m}^{(l)}(\hat s'_l,\hat s_l)
\\ & \times
\bigg\{
v_{l}(\{\hat p, \hat f\}_{m},\hat s_{m},\hat s_{l},s_l)\,
v_{l}^*(\{\hat p, \hat f\}_{m},\hat s'_{m},\hat s'_{l},s'_l)
\\ & -
4\pi \alpha_{\rm s}\
\delta_{s_l \hat s_l}\, \delta_{s'_l \hat s'_l}\
\frac{
{\varepsilon(\hat s_m)^* \!\cdot\!\hat p_l}\
{\varepsilon(\hat s'_m) \!\cdot \!\hat p_l}
}
{(\hat p_{m}\!\cdot\!\hat p_l)^2}
\\& + 
\theta(l\in \{1,\dots,m-1\}, \hat f_l = \hat f_{m} = {\rm g})
\\ & \times 
\left[
v_{2,l}(\{\hat p, \hat f\}_{m},\hat s_{m},\hat s_{l},s_l)\,
v_{2,l}^*(\{\hat p, \hat f\}_{m},\hat s'_{m},\hat s'_{l},s'_l)
\right.
\\ & \quad - 
\left.
v_{3,l}(\{\hat p, \hat f\}_{m},\hat s_{m},\hat s_{l},s_l)\,
v_{3,l}^*(\{\hat p, \hat f\}_{m},\hat s'_{m},\hat s'_{l},s'_l)
\right]
\\ & +
4\pi \alpha_{\rm s}
\
\delta_{s_l \hat s_l}\, \delta_{s'_l \hat s'_l}
\Big[\
\frac{
{\varepsilon(\hat s_m)^* \!\cdot\!\hat p_l}\
{\varepsilon(\hat s'_m) \!\cdot \!\hat p_l}
}
{(\hat p_{m}\!\cdot\!\hat p_l)^2}
\\ & - 
 A_{lk}(\{\hat p\}_{m},\hat s_m',\hat s_m)
\frac{
{\varepsilon(\hat s_m)^* \!\cdot\!\hat p_l}\
{\varepsilon(\hat s'_m) \!\cdot \!\hat p_k}
+{\varepsilon(\hat s_m)^* \!\cdot\!\hat p_k}\
{\varepsilon(\hat s'_m) \!\cdot \!\hat p_l}
}
{\hat p_{m}\!\cdot\!\hat p_l\ \hat p_{m}\!\cdot\!\hat p_k}
\Big]
\bigg\}
\;\;.
\end{split}
\end{equation}
Here we have adopted the shorthand notation
\begin{equation}
\varepsilon(\hat s_m)^* = \varepsilon(\hat p_{m}, \hat s_m;\hat Q)^*
\;\;,
\qquad
\varepsilon(\hat s'_m) = \varepsilon(\hat p_{m}, \hat s'_m;\hat Q)
\;\;.
\end{equation}

One can set
\begin{equation}
\begin{split}
\label{eq:Alkspindependent}
A_{lk}(\{\hat p\}_{m}, \hat s'_m, \hat s_m) 
={}& 
\frac{\hat p_{m}\!\cdot\!\hat p_l\ \hat p_{m}\!\cdot\!\hat p_k}
{\varepsilon(\hat s_m)^* \!\cdot\!\hat p_l\
\varepsilon(\hat s'_m)\!\cdot \!\hat p_k
+\varepsilon(\hat s_m)^* \!\cdot\!\hat p_k\
\varepsilon(\hat s'_m)\!\cdot \!\hat p_l
}
\\
&\times
\left(
\frac{\varepsilon(\hat s_m)^* \!\cdot\!\hat p_l\
\varepsilon(\hat s'_m)\!\cdot \!\hat p_l}{(\hat p_{m}\!\cdot\!\hat p_l)^{2}}
-A_{lk}'(\{p\}_m)\,
\frac{\varepsilon(\hat s_m)^* \!\cdot\!\hat P_{lk}\
\varepsilon(\hat s'_m)\!\cdot \!\hat P_{lk}}
{(\hat p_{m}\!\cdot\!\hat p_l\ \hat p_{m}\!\cdot\!\hat p_k)^{2}}
\right)\;\;.
\end{split}
\end{equation}
The matrix $A_{lk}'$ should be positive and obey $A_{lk}'+ A_{kl}' = 1$. Thus one can take one of the choices given above for it. Then
\begin{equation}
A_{lk}(\{\hat p\}_{m}, \hat s'_m,\hat s_m) 
+ A_{kl}(\{\hat p\}_{m}, \hat s'_m, \hat s_m) = 1
\;\;.
\end{equation}

With this form for $A_{lk}$, the last term in Eq.~(\ref{eq:newspinglue2}) simplifies so that we obtain
\begin{equation}
\begin{split}
\label{eq:newspinglue3}
y_{m-1}^{(l)}(s'_l,s_l)
={}&
\frac{C_{\rm F}\,S_l(\{\hat f\}_{m})}{\Phi_{lk}(\{\hat p,\hat f\}_{m})}
\sum_{\hat s'_{m},\hat s_{m}}
y_{m}^{(m)}(\hat s'_{m},\hat s_{m})\,
\sum_{\hat s'_l,\hat s_l}
y_{m}^{(l)}(\hat s'_l,\hat s_l)
\\ & \times
\bigg\{
v_{l}(\{\hat p, \hat f\}_{m},\hat s_{m},\hat s_{l},s_l)\,
v_{l}^*(\{\hat p, \hat f\}_{m},\hat s'_{m},\hat s'_{l},s'_l)
\\ & -
4\pi \alpha_{\rm s}\
\delta_{s_l \hat s_l}\, \delta_{s'_l \hat s'_l}\
\frac{
{\varepsilon(\hat s_m)^* \!\cdot\!\hat p_l}\
{\varepsilon(\hat s'_m) \!\cdot \!\hat p_l}
}
{(\hat p_{m}\!\cdot\!\hat p_l)^2}
\\& + 
\theta(l\in \{1,\dots,m-1\}, \hat f_l = \hat f_{m} = {\rm g})
\\ & \times 
\left[
v_{2,l}(\{\hat p, \hat f\}_{m},\hat s_{m},\hat s_{l},s_l)\,
v_{2,l}^*(\{\hat p, \hat f\}_{m},\hat s'_{m},\hat s'_{l},s'_l)
\right.
\\ & \quad - 
\left.
v_{3,l}(\{\hat p, \hat f\}_{m},\hat s_{m},\hat s_{l},s_l)\,
v_{3,l}^*(\{\hat p, \hat f\}_{m},\hat s'_{m},\hat s'_{l},s'_l)
\right]
\\ & +
4\pi \alpha_{\rm s}
\
\delta_{s_l \hat s_l}\, \delta_{s'_l \hat s'_l}\
A_{lk}'(\{\hat p\}_{m})
\frac{
{\varepsilon(\hat s_m)^* \!\cdot\!\hat P_{lk}}\
{\varepsilon(\hat s'_m) \!\cdot \!\hat P_{lk}}
}
{(\hat p_{m}\!\cdot\!\hat p_l\ \hat p_{m}\!\cdot\!\hat p_k)^2}
\Big]
\bigg\}
\;\;.
\end{split}
\end{equation}
The last term in Eq.~(\ref{eq:newspinglue3}), constructed from the eikonal approximations for the direct and interference graphs, contains the soft and the soft$\times$collinear singularities. The rest of the contributions together represents the pure collinear singularities. In Eq.~(\ref{eq:newspinglue3}), the physics of the soft and soft$\times$collinear term is clear. The polarization of parton $l$ is unchanged and the plane of the decay (as specified by $\hat P_{lk}$) is preferentially aligned  with the polarization of the emitted soft gluon.

We also need the spin averaged splitting function $\Phi_{lk}$ with this choice of $A_{lk}$. Rather than start with its definition, we can use Eq.~(\ref{eq:newspinglue2}). According to Eq.~(\ref{eq:Ynormalization}), if we insert unit matrices for $y_{m}^{(m)}(\hat s'_{m},\hat s_{m})$ and $y_{m}^{(l)}(\hat s'_{l},\hat s_{l})$, the output $y_{m-1}^{(l)}(s'_l,s_l)$ must have trace equal to two. This fixes the factor $\Phi_{lk}$ in Eq.~(\ref{eq:newspinglue2}). This calculation gives Eqs.~(\ref{eq:Philk2}) and (\ref{eq:Alkresult}) for $\Phi_{lk}$ with the factor $A'_{lk}$ that was used for the spin dependent $A_{lk}$, Eq.~(\ref{eq:Alkspindependent}).

It is of interest to examine the soft gluon spin factor from Eq.~(\ref{eq:newspinglue3}),
\begin{equation}
F(\hat s'_m,\hat s_m) =
\frac{
{\varepsilon(\hat p_{m}, \hat s_m;\hat Q)^* \!\cdot\!\hat P_{lk}}\
{\varepsilon(\hat p_{m}, \hat s'_m;\hat Q) \!\cdot \!\hat P_{lk}}
}
{-\hat P_{lk}^2}
\;\;.
\end{equation}
This is normalized to $\sum_s F(s,s) = 1$, so that it represents the contribution to the spin weight function from a single splitting in which the new parton $m$ is a very soft gluon. For this purpose, we assume that partons $l$ and $k$ are parts of a narrow jet with momentum in approximately the $z$-direction in the rest frame of $\hat Q$, the total momentum of all the final state partons. Then we can parameterize $\hat p_l$, $\hat p_k$ and $\hat p_m$ by using two dimensional transverse vectors $\bm{\theta}_l$, $\bm{\theta}_k$ and $\bm{\theta}_m$ as, with the notation $p = (p^0,\bm p,p^3)$,
\begin{equation}
\begin{split}
p_l ={}& E_l\,(1,\bm{\theta}_l, \sqrt{1 - \bm{\theta}_l^2})
\;\;,
\\
p_k ={}& E_k\,(1,\bm{\theta}_k, \sqrt{1 - \bm{\theta}_k^2})
\;\;,
\\
p_m ={}& E_m\,(1,\bm{\theta}_m, \sqrt{1 - \bm{\theta}_m^2})
\;\;.
\end{split}
\end{equation}
We will assume that $|\bm{\theta}_l| \ll 1$ and $|\bm{\theta}_k| \ll 1$. Then, because of destructive interference, the important integration region for $\bm{\theta}_m$ is $|\bm{\theta}_m| \ll 1$. We can then make small angle approximations everywhere. For the polarization vectors we can approximate
\begin{equation}
\varepsilon \approx (0,\bm{\varepsilon},- \bm{\theta}_m \cdot \bm{\varepsilon})
\end{equation}
with $|\bm{\varepsilon}|\approx 1$. Then $p_m\cdot\varepsilon \approx 0$. With these approximations
\begin{equation}
\begin{split}
F(&\hat s'_m,\hat s_m) \approx
\\&
\frac
{
[(\bm{\theta}_l - \bm{\theta}_m)^2 \,
(\bm{\theta}_k - \bm{\theta}_m)\cdot \bm{\varepsilon}(\hat p_{m}, s_m;\hat Q)^*
-(\bm{\theta}_k - \bm{\theta}_m)^2 \,
(\bm{\theta}_l - \bm{\theta}_m)\cdot \bm{\varepsilon}(\hat p_{m}, s'_m;\hat Q)
]^2
}
{(\bm{\theta}_l - \bm{\theta}_m)^2
(\bm{\theta}_k - \bm{\theta}_m)^2
(\bm{\theta}_l - \bm{\theta}_k)^2}
\;\;.
\end{split}
\end{equation}

We can combine $F(s',s)$ with the spin decay matrix for parton $m$ in the form of Eq.~(\ref{eq:ydecomposition})
\begin{equation}
\label{eq:ydecomposition2}
y_m^{(m)}(\hat s'_m,\hat s_m) = \hat Y_m\left[
2\hat P_m\, \hat \xi_m(\hat s'_m)\, \hat \xi_m^*(\hat s_m) 
+ (1-\hat P_m)\,
\delta_{\hat s'_m,\hat s_m}
\right]
\;\;.
\end{equation}
In Fig.~\ref{fig:softdist}, we display
\begin{equation}
\label{eq:softspinfactor}
\langle F \rangle = \sum_{s'_m,s_m}
y_m^{(m)}(\hat s'_m,\hat s_m)
F(\hat s'_m,\hat s_m)
\end{equation}
as a function of the angles $(\theta_{m,x},\theta_{m,y})$ of the soft gluon $m$. We have taken $\bm{\theta}_l = (0.1,0)$ and $\bm{\theta}_k = (-0.1,0)$.
We have set $\hat Y_m = 1$ with polarization $\hat P_m = 0.15$ in the direction
\begin{equation}
\sum_s \hat \xi_m(\hat s'_m)\bm{\varepsilon}(\hat p_{m}, \hat s'_m;\hat Q)
= (1,0)
\;\;.
\end{equation}
We see that for this polarization choice, the spin factor gives positive interference for $\bm{\theta}_m$ in the region between $\bm{\theta}_l$ and $\bm{\theta}_k$. This is in addition to the positive interference in the spin-averaged cross section in this region that we took note of in Ref.~\cite{NSspinless}. For polarization in the $(0,1)$ direction, one has the opposite effect, $\langle F \rangle_{(0,1)} = 1 - \langle F \rangle_{(1,0)}$.

\FIGURE{
\centerline{\includegraphics[width = 10 cm]{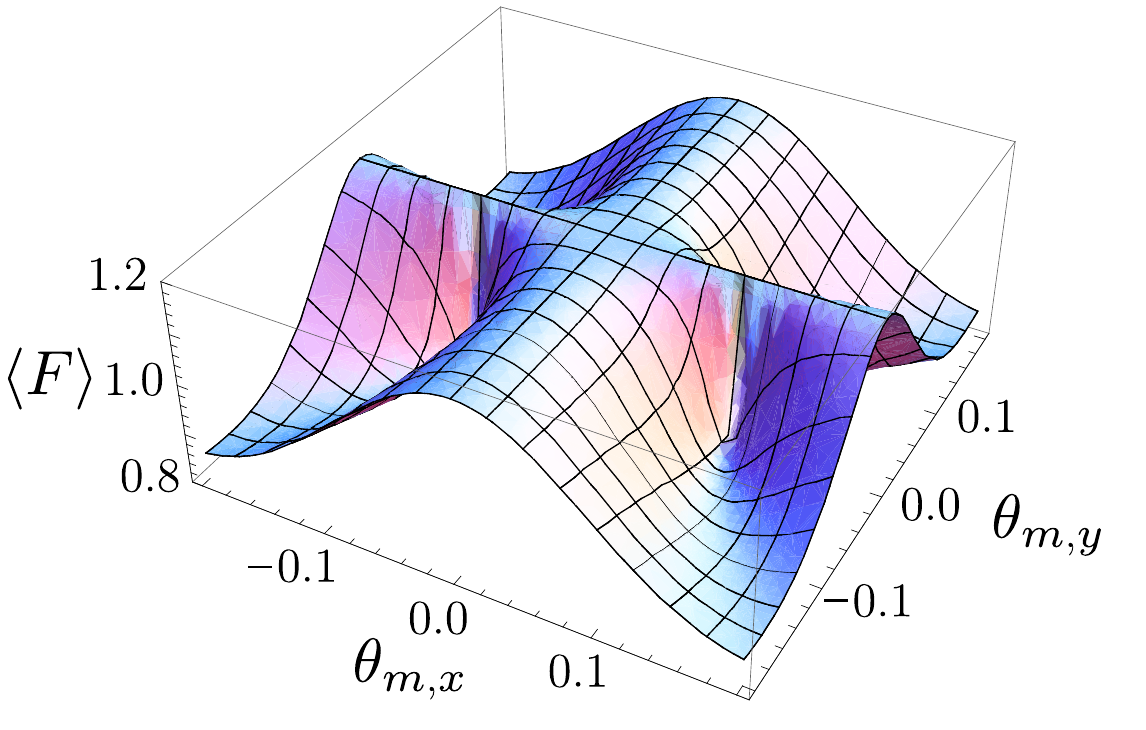}}
\caption{The soft gluon spin weight function $\langle F \rangle$ from Eq.~(\ref{eq:softspinfactor}), plotted against the emission angle $(\theta_{m,x},\theta_{m,y})$ of the soft gluon. Parton $l$ is at $\bm{\theta}_l = (0.1,0)$ and parton $k$ is at $\bm{\theta}_k = (-0.1,0)$. The soft gluon is assumed to be 15\% polarized in the $(1,0)$ direction.}
\label{fig:softdist}
}

\section{Conclusions}
\label{sec:conclusions}
 
In Ref.~\cite{NSspinless}, we have seen how to formulate a leading color, spin averaged shower version of the general parton shower formalism of Ref.~\cite{NSshower} in a fashion that could be implemented in a standard style of calculation as a Markov process. However, this approximation leaves out potentially important effects related to spin and color. In this paper, we have seen how one might add back the effect on the parton momentum distributions of the parton spin correlations, while still working in the leading color approximation.

We assume that the spins of the partons that appear at the end of the shower are not measured by the measurement function applied to the state produced after hadronization of these partons. Then the probability to get a given shower history and set of parton momenta at the end of the shower is the same probability as for a spin averaged shower but times a certain spin weight function. We have seen that this spin weight function can be obtained from a straightforward computation that takes an amount of computer time that is linear in the number of partons. 

If a parton shower event generator produces events with weights $w_i$ instead of events with weight 1, there is a potential to degrade the numerical convergence rate of the calculation of a desired cross section: it may take more Monte Carlo events to produce a result with the same statistical accuracy. This can happen if the weights have a large variance. For instance, suppose that for a certain observable of interest the weights average to $\bar w = 0.6$, indicating that the expectation value of this observable is only 60\% as large as it would be for a spin-averaged shower. If almost all of the weights lie between $-1\times \bar w$ and $3\times \bar w$, there is no real loss of accuracy, but if the weights are typically spread over the range $-100\times\bar w < w_i < + 100\times\bar w$, then many more Monte Carlo events will be needed to produce the same accuracy as with all equal weights. In this event, the simplest approach would be to calculate the spin weight factor corresponding to just the first $N$ splittings after the hard interaction. The true answer is then found by taking the limit $N \to \infty$. The maximum value of $N$ that could be used would be limited by the computer power available. It can well be that, for a final state observable of interest, $N=10$ is enough. 

On the other hand, if some final state observables are found to be very sensitive to spin, then, for those observables, larger values of $N$ than are practical might needed. In that case, one could modify the algorithm for generating events by choosing azimuthal angles with a probability $\rho' = F\times \rho_0$, where $\rho_0$ is the probability for generating events in the spin averaged case and $F$ is a new factor. For example, one could follow \textsc{Pythia} and make use of the correlations given in Ref.~\cite{WebberCorrelations} between the azimuthal angle of a parton splitting and the azimuthal angle of the splitting that produced the mother parton. Whatever method is used, one wants $F > 1$ for the kinds of events for which the spin weight factor $w_0 \equiv \sbrax{1_{\rm spin}} \sket{\rho^{(N)}_{\rm spin}}$ is larger than 1 and $F<1$ for the kinds of events for which $w_0 <1$.  This would give a new spin weight factor $w'= w_0/F$, so that the new weight factor more nearly approximates 1. It is not necessary to get $F$ to exactly match $w_0$. One simply has to arrange that $w_0/F$ is never very large compared to its average value.

We have given formulas for the calculation of the spin weight function using the splitting functions that appear in the general formulation of Ref.~\cite{NSshower}. However, the basic method is of quite general applicability. To include spin in a parton shower Monte Carlo based on averaging over spins, one needs spin dependent splitting functions with two properties: 1) their soft and collinear limits match QCD matrix elements and 2) their spin averages are the splitting functions used in the existing parton shower Monte Carlo. For instance, one could define spin dependent dipole splitting functions that generalize those of Catani and Seymour \cite{CataniSeymour} for use in a dipole shower \cite{dipoleshowers}. In addition, it should be straightforward to define spin dependent antenna splitting functions \cite{antenna} for use in an antenna shower \cite{Vincia}. With spin dependent splitting functions, one can generate the spin weight functions by working backwards from the final state, as we have seen in this paper.


\acknowledgments{
We thank P.~Skands for helpful conversations. 
This work was supported in part the United States Department of Energy and by the Hungarian Scientific Research Fund grant OTKA T-60432.
}


\end{document}